\documentclass[a4paper,fleqn,usenatbib]{mnras}
\usepackage[T1]{fontenc}
\usepackage{ae,aecompl}
\usepackage{graphicx}
\graphicspath{{images_de_larticle/}}
\usepackage{amssymb}
\usepackage{amsmath}
\usepackage{hyperref}
\usepackage{tabularx}
\usepackage{booktabs}

\usepackage[usenames,dvipsnames]{color}

\newcommand{\jc}[1]{{\bf\textcolor{black}{{#1}}}}

\title[The \texttt{$\pi$-DOC} algorithm]{
Predicting Images for the Dynamics Of stellar Clusters (\texttt{$\pi$-DOC}): a deep learning framework to 
predict mass, distance and age of globular clusters.
}
\author[Chardin \& Bianchini]
{\parbox{\textwidth}{Jonathan Chardin$^{1}$\thanks{E-mail: jonathan.chardin@astro.unistra.fr}
\& Paolo Bianchini$^{1}$
}\vspace{0.4cm}
\\
$^1$Observatoire Astronomique de Strasbourg, Universit\'e de Strasbourg, CNRS UMR 7550, 11 rue de l'Universit\'e, F-67000 Strasbourg, France \\}

\begin{document}
%=========================================================================

%========================================================================= 
%Title Page
%=========================================================================

\date{Accepted / Received }

\pagerange{\pageref{firstpage}--\pageref{lastpage}} \pubyear{2020}

\maketitle

% ABSTRACT ======================================================================
\begin{abstract}
Dynamical mass estimates of simple systems such globular clusters (GCs) still suffer from up to a factor of 2 uncertainty. This is primarily due to the oversimplifications of standard dynamical models that often neglect the effects of the long-term evolution of GCs. Here, we introduce a new approach to measure the dynamical properties of GCs, based on the combination of a deep-learning framework and the large amount of data from direct $N$-body simulations. Our algorithm, \texttt{$\pi$-DOC} (\textit{Predicting Images for the Dynamics Of stellar Clusters}) is composed of two convolutional networks, trained to learn the non-trivial transformation between an observed GC luminosity map and its associated mass distribution, age, and distance. The training set is made of V-band luminosity and mass maps constructed as mock observations from $N$-body simulations. The tests on \texttt{$\pi$-DOC} demonstrate that we can predict the mass distribution with a mean error per pixel of 27\%, and the age and distance with an accuracy of 1.5 Gyr and 6 kpc, respectively. In turn, we recover the shape of the mass-to-light profile and its global value with a mean error of 12\%, which implies that we efficiently trace mass segregation. A preliminary comparison with observations indicates that our algorithm is able to predict the dynamical properties of GCs within the limits of the training set. These encouraging results demonstrate that our deep-learning framework and its forward modelling approach can offer a rapid and adaptable tool competitive with standard dynamical models.

%263

\end{abstract}


%===============================================================================
%Body
%===============================================================================

 %%%%%%%%%%%%%%%%%%%%%%%%%%%%%%%%%%%%%%%%%%%%%%%%%%%%%%%%%%%%%%%%%%%%%%%%%%%
 %
 %                      Introduction  section 1
 %
 %%%%%%%%%%%%%%%%%%%%%%%%%%%%%%%%%%%%%%%%%%%%%%%%%%%%%%%%%%%%%%%%%%%%%%%%%%%%

\section{Introduction}
\label{intro}

One of the key goals of Astrophysics is the understanding of the constituents of the Universe. Amongst others, the mass content of stellar systems is fundamental for unveiling the build-up and origin of galaxies throughout cosmic time, and distinguishing between baryonic and (non-visible) non-baryonic matter.

Globular clusters (GCs) -- being old, gas-free and almost spherical -- are the critical example of stellar systems for which a mass measurement is expected to be straightforward. However, state-of-the-art studies of the dynamics of Milky Way (MW) GCs still suffer from an uncertainty in their mass determination of up to a factor of 2. As an example, even for the extensively studied and nearby GCs $\omega$-Centauri and 47~Tuc, the dynamical mass measurements in the literature range from $\simeq$2 to $5\times10^6 M_\odot$  \citep{Zocchi2017} and from $\simeq$5 to $10\times10^5 M_\odot$ \citep{Bellini2017}, respectively.
This uncertainty, together with the uncertainties due to distance measurements, translates into a factor of 2 uncertainty in the mass-to-light ratio (M/L). This makes a comparison between measured M/L and predicted M/L from stellar population models rather inconclusive and further hinders the search for the possible presence of non-visible matter (e.g. stellar dark remnants or non-baryonic dark matter), which is typically encoded in dynamical M/L discrepancies.

The difficulty in obtaining robust mass determinations resides in GCs rich present-day structures. Some GCs are thought to have accreted onto the MW through merger events (e.g. \citealp{Forbes2010,Massari2019,Myeong2019,Pfeffer2020}), others are known to be the remnant of the nucleus of dwarf galaxies (e.g. M54, \citealp{Ibata1995,Alfaro-Cuello2019}), possibly containing some left over dark matter (e.g. \citealp{Bianchini2019,Wan2021}) and exhibiting clear signs of internal rotation \citep{Kamann2018,Bianchini2018}. Moreover, the Hubble time long interplay between internal processes (stellar evolution and two-body relaxation) and external processes (tidal interaction with the host galaxy) leads to subtle dynamical effects such as the evolution of the stellar mass function (\citealp{Vesperini1997,WebbVesperini2016}), mass segregation and partial energy equipartition (e.g. \citealp{Spitzer1987,Trenti2013,Bianchini2016}), evaporation and formation of tidal tails \citep{Gieles2011,Kuzma2016,Malhan2018}, and pressure anisotropy (\citealp{Baumgardt2003,Watkins2015,Jindal2019}). The main result of these evolutionary ingredients is a significant reshaping of a GC mass profile, leading to a M/L profile that is not constant \citep{Baumgardt2017,Bianchini2017}. On top of this, the presence of stellar remnants (neutron stars and stellar mass black holes; e.g. \citealp{Breen2013,Zocchi2019,Vitral2021}) and possibly intermediate-mass black holes (\citealp{Lutzgendorf2013}) can critically contribute to the complexity of GCs present-day structure. 

A standard technique to measure GCs mass consists in transforming surface brightness profiles into mass density profiles, assuming a constant M/L ratio typical of their stellar population (\citealp{MLvdM2005}). However, this quick technique completely ignores all of the dynamical effects listed above. More sophisticated dynamical modelling (e.g. Jeans models, distribution function models, orbit-superposition models) can account for a number of these ingredients in a simplified way, for example accounting for rotation (\citealp{Varri2012,Bianchini2013}), mass segregation and anisotropy (\citealp{DaCostaFreeman1976,GunnGriffin1979,GielesZocchi2015,Torniamenti2019}), kinematic subcomponents and presence of dark remnants (\citealp{vandeVen2006,denBrok2014,Henault-Brunet2020,Vitral2021}), at the expense of dealing with a significant number of degenerate parameters (e.g. mass-anisotropy degeneracy). Often, these techniques cause biases up to a factor of ~2 in mass estimates (\citealp{Sollima2015,Henault-Brunet2019}). In contrast, direct $N$-body simulations, including stellar evolution and the effects of an external tidal field, naturally take into account all the evolutionary star-by-star effects that shape the mass content of present-day GCs. This comes at an incredibly high computational cost (a realistic simulation can take several months on a GPU machine, \citealp{Heggie2014,Wang2016}), making it unfeasible to directly model MW GCs one-to-one without rescaling techniques (e.g. \citealp{Baumgardt2017}).

In the light of these limitations, we wish to introduce a new modelling technique capable of 1) measuring GCs dynamical mass, 2) fully incorporating the non-trivial effects of dynamical evolution, and 3) building on the recent developments of machine and deep learning techniques. Deep learning techniques have become widely used to make automatic predictions in the field of astronomy. They rely on large amount of data to train an algorithm to learn the complex relationships -- the physics at play -- in a specific astronomical context. In particular, the increasing amount of data coming from large simulations provide an ideal data base on which apply and exploit the advantages of such techniques. 

In the field of stellar clusters dynamics, machine learning techniques have been applied to a limited number of studies (e.g., \citealp{Pasquato2016,Askar2019,Bialopetravicius2019}).
%with either the goal of discriminating between formation mechanisms (monolithic collapse vs. mergers, PasquatoChung2016) or identifying clusters with possible presence of black holes subsystems (Askar2018).
However, the growing number of GCs dynamical simulations available to the community makes the applicability of machine learning currently feasible. A typical GC simulation, such as a direct $N$-body simulation of a 13 Gyr old GC, contains $\sim10^3$ time snapshots (e.g. \citealp{Bianchini2018b}). For each of these snapshots, there is a unique physical connection between observable quantities (such as surface brightness and velocity dispersion), and intrinsic quantities, namely the mass distribution, age and distance of the cluster. Moreover, all the complex dynamical mechanisms and physical processes responsible for GCs evolution are encoded self-consistently. Their highly unexploited predictive power makes GC $N$-body simulations the ideal data base to construct an extensive training set suitable for deep learning. 

Our goal is to develop an algorithm, \texttt{$\pi$-DOC} (\textit{Predicting Images for the Dynamics Of stellar Clusters}), capable of learning the non-trivial transformation between observable luminosity maps and the intrinsic mass distribution (and associated M/L), without the need of constructing complex and multi-component dynamical models. Our technique will be directly applicable to observations and includes, in a forward-modelling fashion, a series of observational effects (i.e. seeing conditions, distance) by treating the $N$-body simulations as mock observations. Additionally, the structure of the algorithm will be designed to also provide an estimate of the age and distance of a given GC.

This paper, the first of a series, aims at showing the proof of concept and highlighting the further development needed for an extensive application to observations. The basis of the algorithm -- a convolutional encoder-decoder neural network -- has been developed in \citealp{2019MNRAS.490.1055C} in the context of simulations of reionization and proved to be particularly suitable for inputs such as 2D maps. Here we employ a similar structure in order to predict the 2D mass distribution starting from the luminosity distribution in a given field-of-view (FoV). Moreover, we will implement an additional convolutional neural network (CNN) to predict two scalars (age and distance) starting from the same luminosity maps.

This paper is organized as follows: In Sect. \ref{GCsims}, we detail the $N$-body simulations employed to build the training set as well as the procedure to generate mock GC observations. Sect. \ref{dldetailandtrain} details the neural network framework of \texttt{$\pi$-DOC}, its architecture and the training procedure. In Sect. \ref{resultdect}, leveraging on mock observations, we quantify the performance of the different neural networks. A preliminary application to a set of observations is performed in Section \ref{obs}. We then discuss the success and limitation of \texttt{$\pi$-DOC} in Sect. \ref{discusect} before reporting our conclusions and planned improvements for the near future.

  %%%%%%%%%%%%%%%%%%%%%%%%%%%%%%%%%%%%%%%%%%%%%%%%%%%%%%%%%%%%%%%%%%%%%%%%%%%
 %
 %                      Simulation details section 2
 %
 %%%%%%%%%%%%%%%%%%%%%%%%%%%%%%%%%%%%%%%%%%%%%%%%%%%%%%%%%%%%%%%%%%%%%%%%%%%%

%--------------------------------------%
\begin{table*}

\begin{tabular}{lccccccccc}    \toprule
Simulation &  half-mass radius & distances & PSF & FoV & image size & Training set & Validation set & Test set\\
  & (pc) &  (kpc) & (arcsec) & (arcsec$^2$) &  (pixels) &  &  &  \\\midrule

 MW15-R1.6 &   1.6  & 10-25-50-75-100  & 0.5 - 1.5  	& 40$\times$40 & 160$\times$160  & 10000  & 1000  & 0 \\ 
 MW15    & 3.2 &15-20-60-80            & 1.0               & 40$\times$40 & 160$\times$160  & 0  & 0  & 8172 \\ 
 MW15-R4  & 4.0 & 10-25-50-75-100    & 0.5 - 1.5 	& 40$\times$40 & 160$\times$160  & 10000  & 1000  & 0  \\\bottomrule
 \hline
\end{tabular}
 \caption{Properties of the different simulations used to train, validate and test the different neural networks.
 For each simulation, we give the distance and PSF we employ to create the mock observations.
 We also list the number of snapshots used for each simulation to build the training, validation and test set, as explained in Section \ref{sec:sets}.}
 
 \label{tab1}
\end{table*}
%--------------------------------------%

\section{Simulations of GCs}
\label{GCsims}

In this section, we describe the set of GC $N$-body simulations employed in this paper and the procedure to generate mock observations, which will be the basis of the training set of our neural networks.

\subsection{$N$-body simulations of GC dynamics}
\label{fNBODY6details}

The GCs simulations used in this paper are run with the direct $N$-body code \texttt{NBODY6tt} (\citealt{2011MNRAS.418..759R}; \citealt{2015MNRAS.448.3416R}). This code is an extension of \texttt{NBODY6} (\citealt{2003gnbs.book.....A}; \citealt{2012MNRAS.424..545N}), which allows for a dynamical evolution of a star cluster in an arbitrary time-dependent tidal field, and incorporates stellar evolution and the presence of a stellar mass function.
We specifically use three simulations originally presented in \cite{2016MNRAS.456..240M} and \cite{2017MNRAS.471.1181B}, all characterized by the same tidal field, namely a circular orbit around the MW centre at a distance of 15 kpc. The MW is modelled as a point mass bulge, plus a disc and a logarithmic halo, as described in \cite{Miholics2014}.

The initial conditions are drawn from a \cite{Plummer1911} sphere with N=50,000 stars and initial half-mass radius of 1.6, 3.2, and 4 pc, for simulation MW15-R1.6, MW15, and MW15-R4, respectively (see Table \ref{tab1}). The initial stellar mass function is a \citet{Kroupa2001} mass function with lower and upper mass limits of 0.1 and 50 $M_{\odot}$, and a metallicity of z=0.001 ([Fe/H]=-1.3). The stars are evolved according to the stellar evolution prescriptions implemented in the code, without considering primordial binaries. All simulations are evolved until 14 Gyr. 

For each simulation we have over 1000 time-snapshots (5781, 2044, 1462, for MW15-R1.6, MW15, and MW15-R4, respectively) each containing the info on 3D position, velocity, stellar mass, stellar radius and bolometric luminosity of each star. We calculate the V-band magnitudes of each stars using a bolometric correction from \citealp{Reed1998} as in \citealp{Heggie2014}.\footnote{As noted in \citealp{Giersz2003}, the effective temperature $T$ in eq. 5 of \citealp{Reed1998} should be replaced by $10^{-4}T$.}

This set of simulations provide a realistic standard representation of the long-term dynamical evolution of a GC, despite the relatively small number of stars employed. Indeed, the three different initial conditions allow us to explore a wide range of dynamical effects that strongly depend on the initial mass density of a cluster, such as mass loss, mass segregation, and velocity anisotropy. At this stage, this simple configuration is ideal to assess straightforwardly the performances of our neural networks and the connection with the physics underneath it. 

In the next section we describe the data treatment procedure to construct the luminosity and mass maps that will be the basis for our training.

\subsection{Generating mock observations of GCs}
\label{fakeGCobs}

From the three simulations described in the previous section, we generate mock observations in post processing. The goal is to construct realistic luminosity maps with all the significant observational effects taken into consideration.  

In order to do so, we exploit the code \texttt{SISCO} (\citealp{Bianchini2015}) originally developed to create mock observations of integrated-light kinematics of GCs starting from a snapshot of a $N$-body simulation (\citealp{deVita2017,Askar2017}). We adapted the code in order to only retrieve the brightness information of a given simulation after defining the observational specification (see Table \ref{tab1}). In particular, we choose a FoV of 40$\times$40 arcsec$^2$, with a pixel size of 0.25 arcsec, giving a total image of 160$\times$160 pixels.\footnote{While a pixel scale of 0.25 arcsec can be considered as typical of ground-based instruments, the actual size of the FoV selected here is relatively small; this is to avoid memory limitations at this phase of development.} Moreover, we define a point spread function (PSF) with a Moffat distribution (\citealp{Moffat1969}, with shape parameter $\beta=2.5$ \citealp{Trujillo2001}) allowing to describe more extended PSF wings that a standard Gaussian PSF. Given the PSF shape, we further specify the seeing conditions of a mock observation (i.e. the FWHM of the PSF) as 0.5, 1.0 and 1.5 arcsec (typical values of ground based observations).

We additionally assign the distance of a cluster varying from 10 to 100 kpc (see Section \ref{sec:sets}), allowing us to explore a large range of different surface brightness densities. These specifications are chosen to be representative of a generic enough case, while carefully including all the most relevant observational effects. Finally, for each snapshot, we decide to implement a random shift of the GC centre from 0 to $\pm5$~arcsec (corresponding to a maximum shift of $\pm20$ pixels) and a random rotation of either 0, 90, 180, or 270$^{\circ}$. This is done to account for the possibility of off-centre GC observations and avoid that a neural network algorithm could only learn to predict mass distributions that are perfectly centred or oriented in a particular direction. We notice that at this stage we do not take into account the effects of background contamination. This effect is expected to be negligible for a standard GC with high stellar density in its central and intermediate regions, which can reach up to $10^5$ M$_\odot$ pc$^{-3}$. However, this effect could become relevant for low-density stellar clusters in dense background regions (see e.g. \citealp{Bialopetravicius2020}).

For each snapshots we produce a flux map in the V-band and an associated intrinsic mass map, with the same FoV and pixels scale, but without considering the PSF. After producing the mock observations, we additionally smooth the flux/mass maps (see e.g. Figure \ref{Autoencoder_archi}). This allow us to smooth regions of sharp transitions between individual stars and assure better performances in the training of our convolutional neural networks, which are known to underperform with discrete maps.
In our case, we choose to Gaussian smooth the maps in both directions with a standard deviation for the Gaussian kernel equal to 4 pixels which corresponds to 1 arcsec in our mock observations. After doing this, we checked that our global properties in each maps are conserved (e.g. the total luminosity and the total mass). 

Our sample covers a total mass in the FoV ranging between $\sim 10^3$ and $\sim 10^{4.5} \, \mathrm{M_{\odot}}$. These maps are the basis for the training, validation, and test sets, as described in the following section. We remind that for each flux/mass map pair considered here, there is also an associated GC distance and age.

\subsection{The training, validation, and test sets}
\label{sec:sets}

When training a neural network, we first need a \textit{training set}, which is composed of the maps that are actually exploited during the training phase. Secondly, we need a \textit{validation set}, which is used to monitor how the current trained version of the network performs at predicting maps that are not part of the training.

In our case, both the training and the validation sets are built from the MW15-R1.6 and the MW15-R4 simulations, which have different initial conditions and therefore represent GCs with different dynamical evolutions. As summarized in Table \ref{tab1}, for these two simulations we consider distances of 10, 25, 50, 75, 100 kpc and seeing conditions of 0.5 and 1.5 arcsec. Therefore, for each simulation we have 10 times the original number of snapshots to build the training set and the validation set. This corresponds to $\mathrm{10 \times 5781 = 57810}$ snapshots for the MW15-R1.6 simulation and $\mathrm{10 \times 1462 = 14620}$ snapshots for the MW15-R4 simulation. 

Since we want to build homogenous training and validation sets, we choose to keep the same number of snapshots in both simulations. Therefore, we randomly select 10000 snapshots in each simulation to build the training set which finally contains 20000 luminosity/mass map pairs. Moreover, we pick 1000 maps from each of the two simulations to build the validation set which consist of a total of 2000 snapshots. When building the validation set we make sure to not pick those maps that are included in the training set.

Thirdly, after finishing a training phase, we need a \textit{test set}, which is composed of maps not used for neither the training nor the validation phase and that come from a different simulation. In our case, we select maps from the MW15 simulation, which have initial conditions (i.e. half-mass radius) and mock observation conditions (i.e. distance and seeing conditions) that are intermediate with respect to the other two simulations (see Table \ref{tab1}). In particular, we select distances of 15, 30, 60, 80 kpc and seeing condition of 1 arcsec. The MW15 simulation is discretized in 2043 snapshots covering the 14 Gyr evolution of the GC.
We choose to keep all of these maps to test our algorithm, for a total number of maps of $\mathrm{4 \times 2043 = 8172}$ when considering the four different distances.
The test set allows us to put at test the neural network algorithm, ensuring that 1) it is able to generalize well over a large set of conditions never seen during the training phase, and that 2) it does not over-learn particular features from the particular simulations used during the training.
Having built these different sets, we now have to settle the architecture of the neural networks by trials and errors to obtain the best performances, as detailed in next section (Sect. \ref{dldetailandtrain}).

 %%%%%%%%%%%%%%%%%%%%%%%%%%%%%%%%%%%%%%%%%%%%%%%%%%%%%%%%%%%%%%%%%%%%%%%%%%%
 %
 %                      Neural net arch and training section 3
 %
 %%%%%%%%%%%%%%%%%%%%%%%%%%%%%%%%%%%%%%%%%%%%%%%%%%%%%%%%%%%%%%%%%%%%%%%%%%%%

\section{Designing and training \texttt{$\pi$-DOC}}
\label{dldetailandtrain}

 \begin{figure*}
   \begin{center}
      \includegraphics[width=\textwidth]{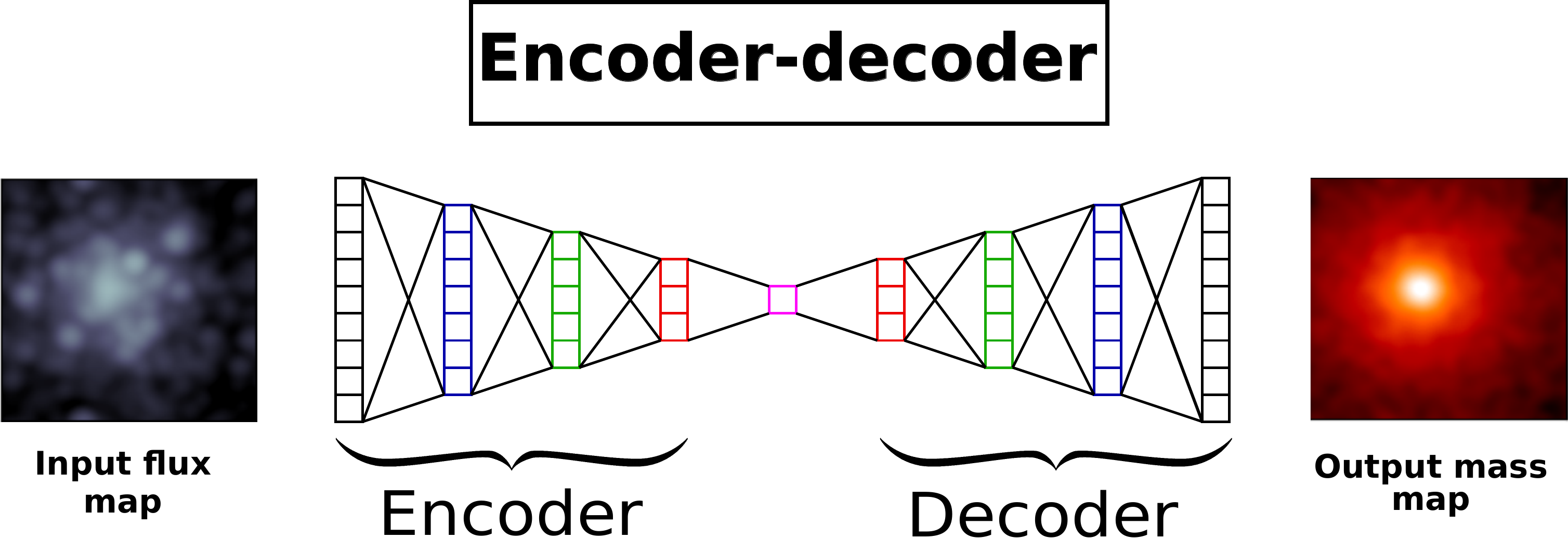}   
  \caption{Architecture of the convolutional encoder-decoder part of \texttt{$\pi$-DOC}.
  This first part of the network takes a GC flux map as an input and gives a mass map as an output.
  Images have 160$\times$160 pixels sizes both for the input and the output. 
  The encoder (decoder) are built with four convolutions (deconvolutions) hidden layers.}
    \label{Autoencoder_archi}
  \end{center}
 \end{figure*}

\begin{figure*}
   \begin{center}
      \includegraphics[width=\textwidth]{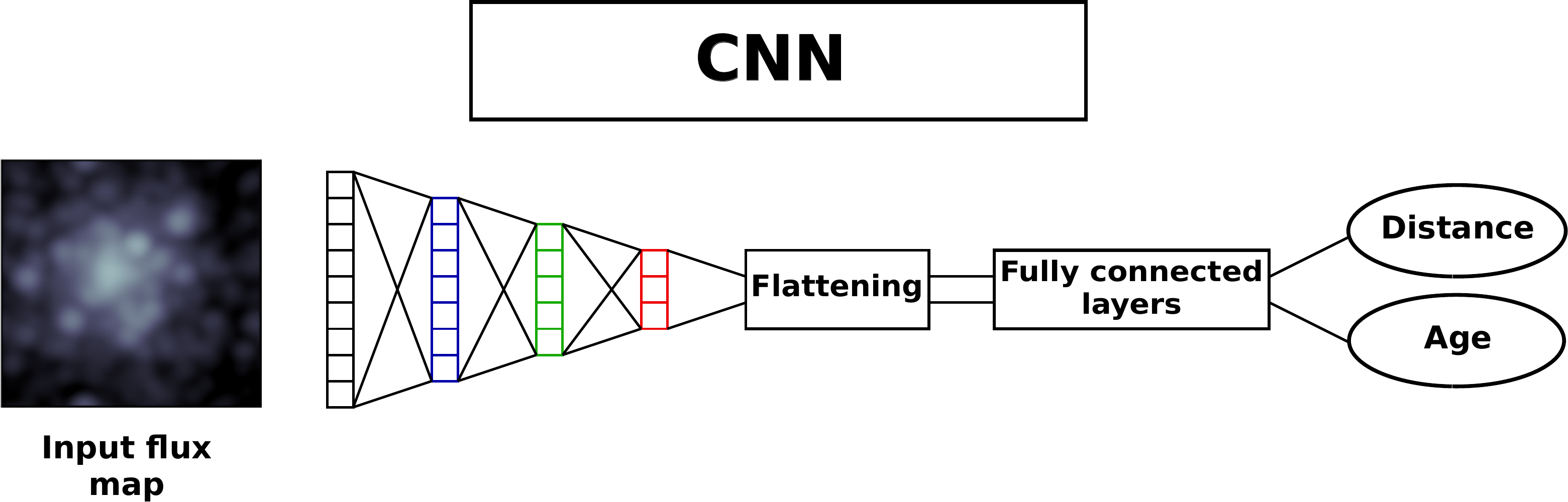}   
  \caption{Architecture of the \texttt{$\pi$-DOC} convolutional neural network (CNN) to predict the age and distance of clusters from their observed luminosity maps.}
    \label{CNNdistage_archi}
  \end{center}
 \end{figure*}

In this section we explain how we use our set of simulations to train a deep learning model to
predict the mass distribution, distance and age of a GC, given its flux map.
We first describe the architecture of the two neural networks constituting our full deep learning framework, 
before showing the training procedure and the performance we can reach.

\subsection{\texttt{$\pi$-DOC} architecture}
\label{PIDOCarch}

The architecture of \texttt{$\pi$-DOC} is composed of two distinct neural networks.
The first network is a convolutional encoder-decoder neural network predicting the mass distribution of a GC in a given FoV, while the second one is a classical convolutional network designed to predict the distance and the age.

\subsubsection{Convolutional encoder-decoder for mass distribution prediction}
\label{autoencoderarchi}

A convolutional encoder-decoder neural network is a deep learning based architecture that aims at predicting a full physical field (i.e. mass map) starting from another field taken as an input (i.e. flux map). In our specific case, the size of the input map and the one of the output map are the same (160$\times$160 pixels).
The architecture, divided into encoder and decoder, is similar (but not exactly the same) to the well known U-net architecture \citet{Unetciataion2015} used for biomedical image segmentation.
The network is composed of a succession of convolution (encoder part) + deconvolution layers (decoder part). 
The encoder and the decoder have the same number of layers in such a way that the decoder is the symmetrical part of the encoder.
Networks of this type have already been used in astronomy to predict physical fields (see e.g. \citealt{2019MNRAS.490.1055C}, \citealt{2020arXiv200614305}, \citealt{2020arXiv200707267} and \citealt{2020arXiv200710340}).

In Fig. \ref{Autoencoder_archi}, we report a scheme of such architecture for the specific case of \texttt{$\pi$-DOC}. Precisely, our encoder/decoder is composed of four convolutional + deconvolution layers.
We use the classic Adam algorithm as an optimizer and a learning rate of $10^{-2.5}$ after some trials.
The loss function chosen to minimize the error between real values and predictions is the perceptual loss (see for example \citealt{Johnson2016Perceptual}) which is detailed in Appendix \ref{perceploss}. In a few words, instead of a per-pixel loss between the output and ground-truth images, we define and optimize a loss function based on high-level features extracted from a pretrained neural network. This procedure has been shown to improve the results in high-quality image generations for image transformation problems, where an input image is transformed into an output image (see \citealt{Simonyan2013}, \citealt{Szegedy2014}, \citealt{Yosinski2015} and \citealt{Nguyen2016}).
In order to avoid overfitting, we also add a dropout regularization layer after the convolution/deconvolution respectively in the encoder/decoder. 
Dropout regularization is the process of randomly turning off a given number of neurons during the training to prevent the network to overlearn the training set (\citealt{Labach2019SurveyOD}). 
The dropout is coded as a value between zero and one, which is the probability to shut down certain neurons, and we choose a value of $10^{-2.5}$ after trials and errors.
We also add skip connections in the decoder part which is known to improve the performance for such a kind of neural network architecture (\citealt{Oyedotun2020skip}).
Skip connections are a way to add informations at the end of a layer in the decoder before entering the next layer. In our case, we concatenate the current decoded map and the corresponding map with similar size in the encoder part. In practice, the combination of these informations is known to improve the performance of convolutional encoder-decoder.

\begin{figure*}
\begin{center}
    \includegraphics[width=\textwidth]{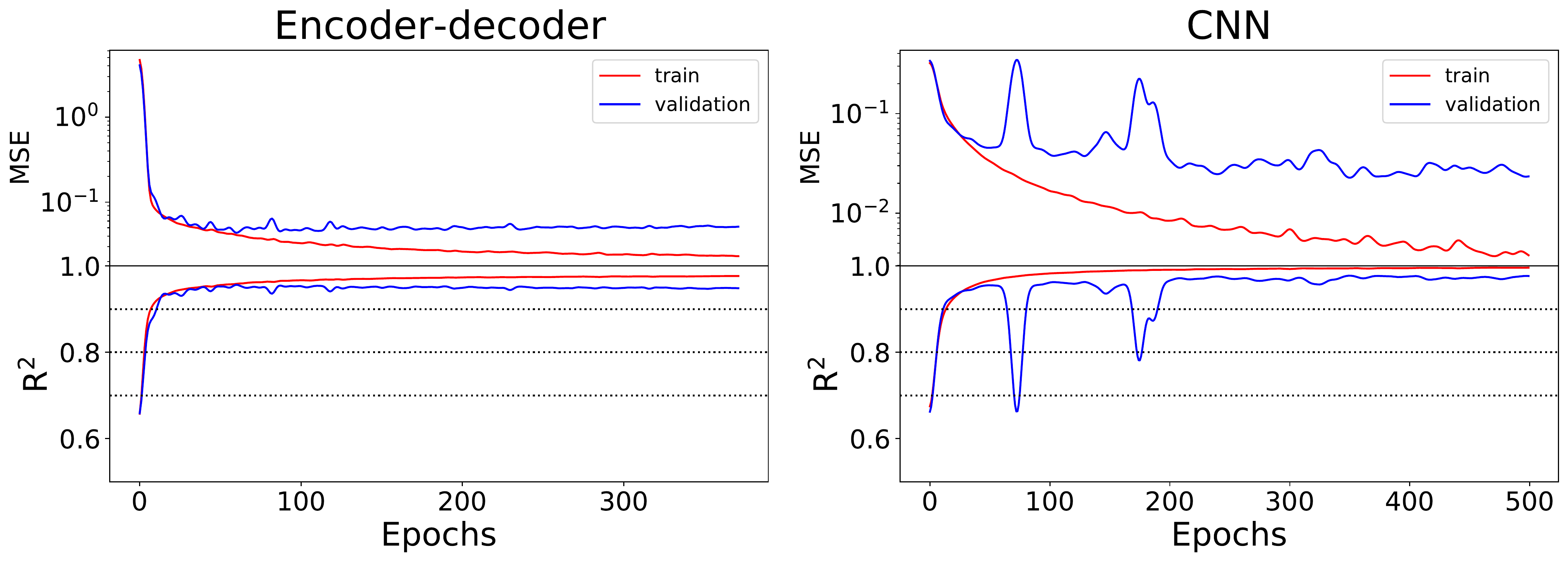}   
\caption{Training performance curves for the whole \texttt{$\pi$-DOC} algorithm. \textit{Left panel}: training curves for the convolutional encoder-decoder for mass map predictions. \textit{Right panel}: training curves for the CNN for ages and distances estimation. In both panels the top curves show the evolution of the mean squared error (MSE) between the predicted values and the real ones as a function of the number of epochs for both the training set and the validation set. The bottom curves show the evolution with the number of epochs of the coefficient of determination $\mathrm{R^2}$ of equation \ref{r2equ}. This allows us to monitor how our model matches the original values and, in particular, how it performs on the unseen data of the validation set. The different horizontal dashed lines show values of $\mathrm{R^2}$= 0.7, 0.8 and 0.9.}
\label{traininggcurve}
\end{center}
\end{figure*}

 \subsubsection{CNN for GC's distances and ages estimation}
\label{agedistarchi}

The second neural network in \texttt{$\pi$-DOC} is a classical convolutional neural network (CNN) similar to very well known neural network classifiers. 
As an alternative to classical artificial neural networks (ANN), CNNs have been shown to be computationally cheaper and have therefore been used extensively in recent years. 
A CNN is composed of a series of successive convolutions and pooling layers (see Fig. \ref{CNNdistage_archi}) before a last output layer.
The last output layer in CNN architecture can take several forms depending on the task the CNN is designed for (i.e. classification, recognition, times series or language processing).

In our case the CNN architecture is built similarly to the CNN designed by \cite{2019MNRAS.484..282G}.
Precisely, our CNN is composed of four convolutional + maxpool layers.
At the end of the last hidden layer, we flatten the data and we add a last fully connected layer with a linear activation function to obtain continuous values (i.e. distance and age) as outputs as opposed to categories in CNN classifiers.
The output of our CNN has the shape of a two-dimensional vector: [$a$, $d$], with $a$ the age and $d$ the distance of the GC. Since it is difficult to train a CNN with values in the range of ages and distances encountered here we normalized these data. We subtract the mean from all the values and we divide by their standard deviation. This ensures that our variables have mean zero and standard deviation of 1.
Each luminosity map is therefore associated with a unique vector of this kind that the CNN is aimed to recover once trained.
As in the convolutional encoder-decoder part of \texttt{$\pi$-DOC}, we use the Adam optimizer with a learning rate of $10^{-4}$ after some trials.
As a loss function, conversely to the encoder-decoder, we use the classic mean square error (MSE) loss. 
We also use dropout regularization to avoid overfitting with a dropout probability of $10^{-2.5}$ after some trials.
 
\subsection{Training the networks}
\label{PIDOCtraining}
 
 \begin{figure*}
   \begin{center}
      \includegraphics[width=\textwidth]{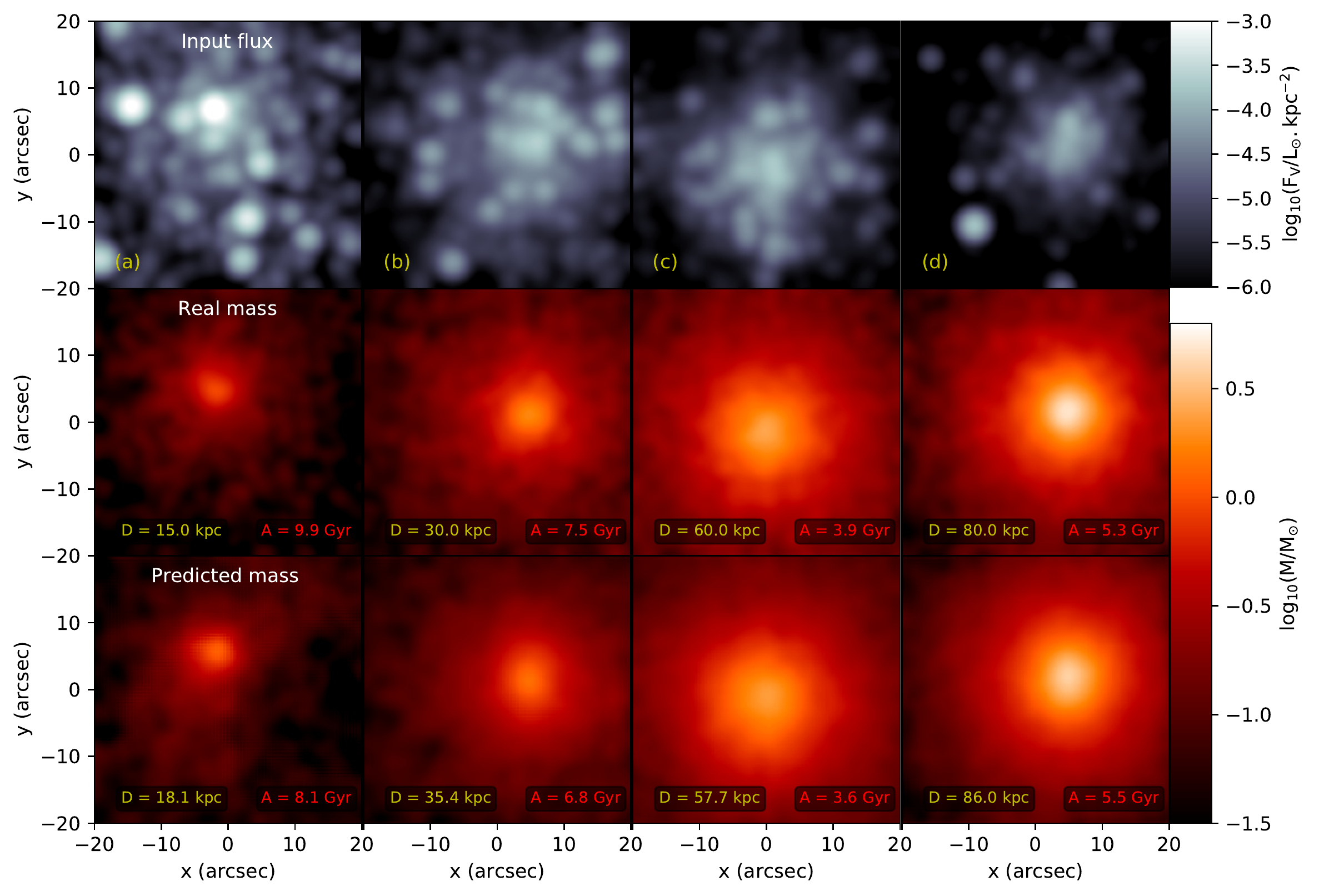}   
  \caption{Full prediction of \texttt{$\pi$-DOC}. First row: example of four input luminosity maps from the original simulation.
  Second row: example of four mass distribution maps from the original simulation corresponding to the luminosity maps from the first row.
  Third row: example of four mass map predicted by \texttt{$\pi$-DOC} by taking the corresponding luminosity maps from the first row as input of the encoder-decoder part. On the different panels of the second and third row, we indicate the distances and ages of the original simulation and the ones predicted by the CNN part of \texttt{$\pi$-DOC}, respectively.}
    \label{predevery}
  \end{center}
 \end{figure*}
 
Both neural networks described in the previous sections have been implemented thanks to the PYTHON API for neural networks Keras (\citealt{2015xxxxcholkeras}). The training of the networks has been done on 4 Tesla V100 GPUs cards on the supercomputer Jean-Zay with HPC resources CINES and IDRIS under the allocation A0070411049 attributed by GENCI. Training the encoder-decoder for mass maps predictions takes about 40 GPUs hours while the CNN part to predict distances and ages takes about 10 hours. Both networks use the same training set composed of 20000 luminosity maps.
 
 To monitor our neural network training, we use two different metrics both for the encoder-decoder and the CNN. 
 The first one is the mean squared error (MSE) between the predicted and the true values (which is actually the loss function used to train the CNN but not the one used for the encoder-decoder).
 The second one is a commonly used indicator called the coefficient of determination $\mathrm{R^2}$ calculated with the following formula (see \citealt{2019MNRAS.484..282G} and \citealt{2019MNRAS.490.1055C}):
 \begin{equation}
 \mathrm{ R^2=\frac{\sum(y_{pred}-\overline{y}_{true})^2}{\sum(y_{true}-\overline{y}_{true})^2}  = 1 - \frac{\sum(y_{pred}-y_{true})^2}{\sum(y_{true}-\overline{y}_{true})^2}} 
\label{r2equ}
 \end{equation}
where $\mathrm{y_{pred}}$ and $\mathrm{y_{true}}$ are the predicted and true values and $\mathrm{\overline{y}_{true}}$ is the average of the true parameters. 
Values of $\mathrm{R^2}$ close to 1 represent a 100\% match between the true data in the original simulation and the ones predicted by the model.

Fig. \ref{traininggcurve} shows the evolution of both indicators for both the convolutional encoder-decoder and the CNN.
The curves shown are the ones corresponding to the best model we kept at the end after trials and errors (i.e. after varying the number of layers, the values of the learning rate, the values of the dropout and by using or not skip connections in both neural networks).
 The upper panel shows the evolution of the MSE for both neural networks as a function of the number of training epochs.
 The red lines show the evolution of the curve for the training set while the blue ones show the same evolution for unseen maps of the validation set.
 
 We observe a quick decrease of the MSE values for both networks for the first tens of epochs.
 After $\sim$ 50 epochs, the validation set begins to saturate at values of about 0.05 for the MSE for the encoder-decoder.
 We report a same kind of `plateau' for the CNN after $\sim$ 200 epochs with MSE $\sim$ 0.03.
 The actual MSE value itself does not give any insight about the actual performance of a given network because it depends on a given deep learning experiment.
 This is why we observe different values between the two networks when we reach the `plateau'.
 From these curves we can only conclude that both networks seem to learn the relation between the input and the output. In particular, they indicate after how many epochs the training is assumed to be achieved: this corresponds to the beginning of the plateau observed for the validation set.

Conversely, the evolution of the $\mathrm{R^2}$ metric can directly be used to compare the training of two distinct networks. 
 In our case, we observe a quick rise of the value of $\mathrm{R^2}$ for both networks with saturating values of $\sim 0.95$ and $\sim 0.98$ for the encoder-decoder and the CNN, respectively, after $\sim$ 100 epochs and $\sim$ 200 epochs, for the validation set (blue curves). 
 These values tell us that both networks can reach very good predictions for unseen maps compared to the ground-truth in the original simulation.
 However, this does not guarantee that our two networks can generalize on new unseen data coming from a different simulation, 
 different from the training and the validation set.
 This is why we have to test our predictions on the test set in order to evaluate the performance of our whole model.
 This is what we extensively test in the next section.

 %%%%%%%%%%%%%%%%%%%%%%%%%%%%%%%%%%%%%%%%%%%%%%%%%%%%%%%%%%%%%%%%%%%%%%%%%%%
 %
 %                      Result section 4
 %
 %%%%%%%%%%%%%%%%%%%%%%%%%%%%%%%%%%%%%%%%%%%%%%%%%%%%%%%%%%%%%%%%%%%%%%%%%%%%
 
 \section{Results}
\label{resultdect}

In this section, we present the tests on the performance of our deep learning model \texttt{$\pi$-DOC}.
We first show how the convolutional encoder-decoder part performs at predicting the mass spatial distribution from the input luminosity map before showing how the CNN part of \texttt{$\pi$-DOC} enables to estimate the distance and age of the GC from the same input maps.
All the results shown from here are done on the test set, i.e. on maps that were never seen during the training phase coming from the MW15 simulation (see Section \ref{sec:sets}). 

\subsection{Mass prediction}
\label{resmass}
We perform different tests to quantify the robustness of our mass map predictor: first, we focus on its ability to generate a 2D mass distribution field, then on the ability to predict radial mass profiles and global mass values (and corresponding M/L) in good agreement with the expected truth.
We use the encoder-decoder to predict all the 8172 mass maps associated to all the flux maps from the test set: 2043 predictions for each of the four distances of the GC from the MW15 simulation that we employ here. The prediction of a single mass map takes about 70 milliseconds on a single Intel Xeon CPU at 2.70GHz. Therefore it takes less than three minutes to predicts the whole 8172 maps of our complete test set.

\subsubsection{Mass field prediction}
\label{resmassdistrib}
In Fig. \ref{predevery}, we show a complete prediction of \texttt{$\pi$-DOC} compared to the ground truth in the original simulation.
The first row shows an example of four flux maps randomly taken at different stages of the evolution of the GC in the test set (i.e. different ages of the GC and different distances). 
These maps are the input of the encoder-decoder part of \texttt{$\pi$-DOC}. The second row shows the corresponding mass maps directly from the original simulation, while the third row shows the prediction from \texttt{$\pi$-DOC}. We stress that all these maps have distances that were never employed during the training (Table \ref{tab1}) and we show here a map for each of the four selected distances. Moreover, the seeing condition (PSF) is of 1.0 arcsec, which is also not part of the training set.

In Fig. \ref{predevery}, the colormaps of the second and third rows are set to the same scale, allowing for a direct comparison.
At first glance we can see that the encoder-decoder predicts maps in good agreement with the ones from the original simulation.
This is true independently from the age of the GC and independently from the concentration of the mass profiles.
Moreover, this holds true whatever the distance, meaning that our neural network is able to generalize well over a wide distance range (we will go back to this in Sect.~\ref{resdistage}).

We can see that the predicted values in each pixel are fully consistent with the real values throughout the extent of the maps, either in the centre and at the outskirts of the GC. This illustrates the ability of the neural network to learn the two-dimensional mass distribution of the GC at various stage of its dynamical life.

\subsubsection{Pixel-by-pixel mass predictions}
\label{truepredres}

\begin{figure}
\begin{center}
    \includegraphics[width=\columnwidth]{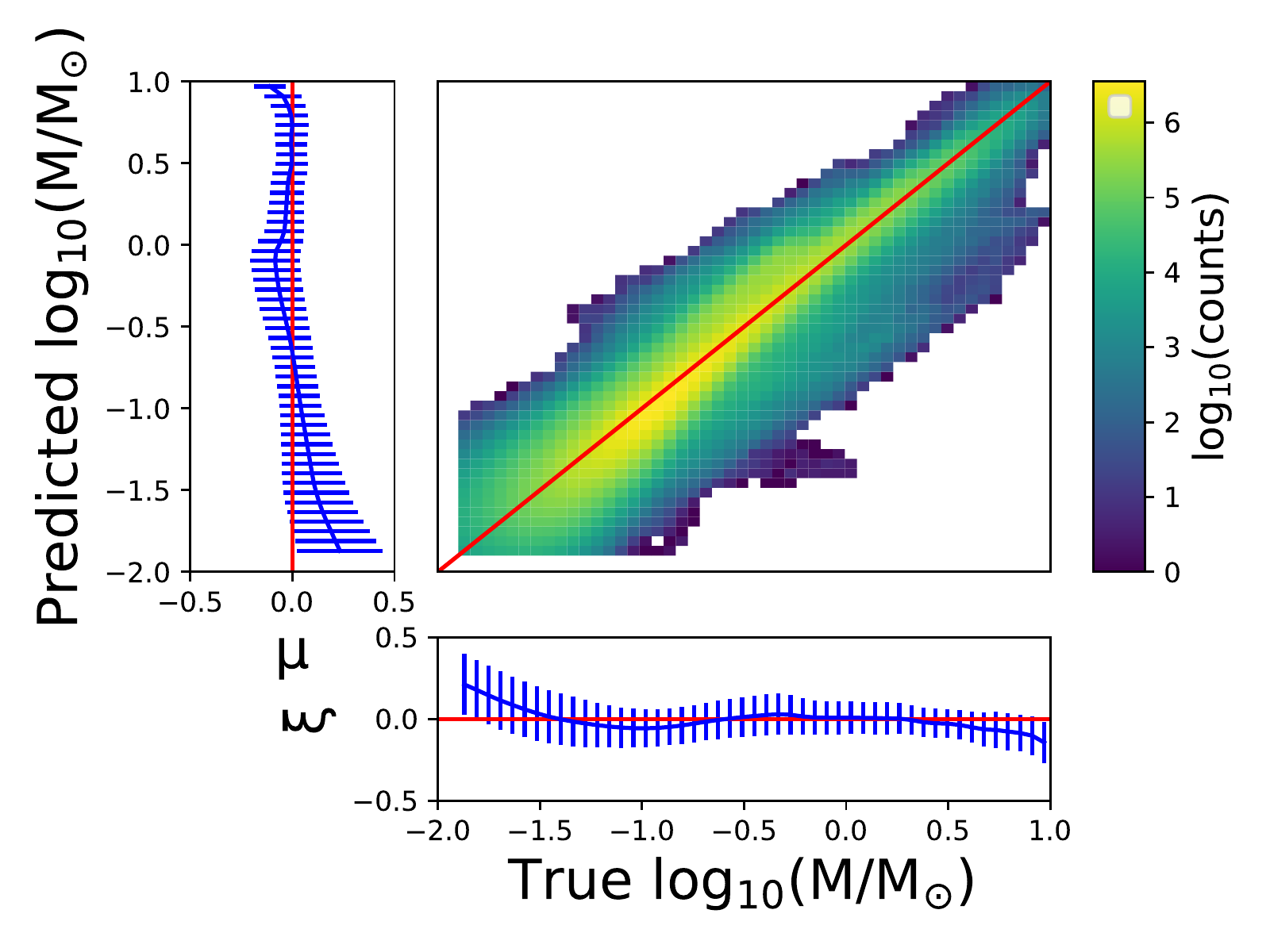}   
\caption{2D histogram of the true versus predicted pixel-by-pixel values of the mass. The histogram is constructed by taking all the pixels over the 8172 mass maps of the test set. The red line shows the one-to-one relation and the colour map encodes the number count of cells lying in the 2D space of true versus predicted values. The bottom and left histograms show the mean and the standard deviation of the residuals $r=M_\mathrm{predicted} - M_\mathrm{true}$ along both axes. The bottom histogram is the learning error, p($r$|True), while the side histogram is the recovery uncertainty, p($r$|Predicted).}
\label{truevspred}
\end{center}
\end{figure}

To go deeply into the analysis, in Fig. \ref{truevspred}, we show the 2D histogram of the true versus predicted mass values for each pixel in all the 8172 maps of the test set. The red line on the figure shows the one-to-one relation. The bottom and left histograms show the mean and the standard deviation of the residuals $r=M_\mathrm{predicted} - M_\mathrm{true}$ along the vertical and horizontal directions, respectively. 

The histogram on the left shows the distribution of residuals as a function of predictions: $\mu$=P(r|Predicted). Here $\mu$ is a measure of the uncertainties on predictions. 
The bottom histogram gives the residuals distribution as a function of true values: $\xi$=p(r|True). Here, $\xi$ is a measure of the encoder-decoder error. On the other hand, $\mu$ is the most relevant quantity to assess the performance of the neural network, since in real life the actual true values are unknown. Therefore we are interested in the range of true values we can reach knowing the prediction of the convolutional encoder-decoder. 

In our case, the encoder-decoder shows a good performance: it predicts values of the mass in each pixel very close to the one-to-one curve.
The average value of the residuals in the uncertainties curve (left side histogram) is fully consistent with the zero residual value for pixels with mass of $\mathrm{log_{10}(M/M_{\odot})}$>\,$-1.5$. 
On the other hand, the encoder-decoder overestimates the smallest values of $\mathrm{log_{10}(M)}$ for $\mathrm{log_{10}(M/M_{\odot})<-1.5}$ and tends to underestimate the largest values for $\mathrm{log_{10}(M/M_{\odot})} \sim 1$. This is actually an inherent feature of neural networks, which appear to have difficulty in predicting the extremum values present in the training set in a deep learning experiment.

To quantify the errors on our predictions in the left side histogram, we calculate for each mass bin the mean percentage error:
\begin{equation}
<\%\, \mathrm{error}>_{\mathrm{bin}} = \frac{100}{\mathrm{N}}\sum{\frac{|M_\mathrm{predicted} - M_\mathrm{true}|}{M_\mathrm{predicted}}}.
\label{percenterror}
\end{equation}
We report an average percentage error of 27.4\% over all mass bins, with a minimum value of 13.7\% and a maximum values of 95.0\%. Notice that the highest percentage errors correspond to the low-values mass bins with $\mathrm{log_{10}(M/M_{\odot})<-1}$. Since these pixels only contain a small fraction of a solar mass, a large error does not imply any significant bias in the prediction on global properties, such as total mass or mass profile, as we will show in Section \ref{massprofandtotmass} and  \ref{massprofandtotmass}. If we exclude these low mass bins, we obtain a mean error of 17.6\%, with a minimum and a maximum of 13.7\% and 22.1\%, respectively, indicating that the encoder-decoder performs well in this higher mass range.

 \begin{figure*}
   \begin{center}
      \includegraphics[width=\textwidth]{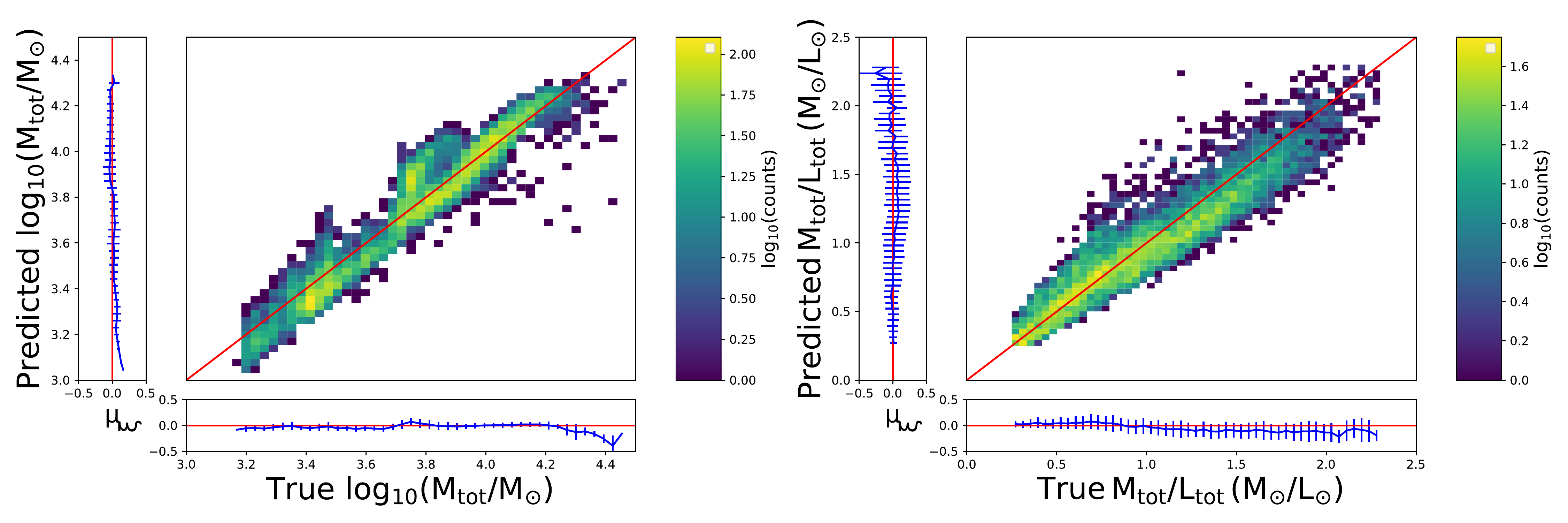}   
  \caption{\textit{Left panel}: 2D histogram of the total true integrated mass versus total predicted mass. The distribution is calculated over the whole test set with 8172 maps with a single integrated value calculated for each map. \textit{Right panel}: Total true integrated mass-to-light ratio versus total predicted mass-to-light ratio. The distribution is calculated over the whole test set with 8172 maps with a single integrated value calculated for each map. In both panels, red lines show the one-to-one relation and colour maps encode the number count in the 2D space of true versus predicted values. The bottom and left histograms show the mean and the standard deviation of the residuals along both axes. Bottom histograms are the learning error, p(r|True), while side histograms are the recovery uncertainty, p(r|Predicted).}
    \label{totmasshisto}
  \end{center}
 \end{figure*}

\subsubsection{Predicted total mass and M/L}
\label{massprofandtotmass}

As a third test, the left panel of Fig. \ref{totmasshisto} shows the 2D histogram of the true total mass $\mathrm{M_{tot}}$ versus the total predicted $\mathrm{M_{tot}}$ for the whole test set. The right panel shows the 2D histogram of the true mass-to-light ratio $\mathrm{M_{tot}/L_{tot}}$ (in V-band) versus the predicted $\mathrm{M_{tot}/L_{tot}}$ also for the whole test set. Notice that to calculate the M/L we use the absolute luminosity assuming that the distance is known.
In both cases, we calculate one of these integrated quantities over one single map and we repeat this for all the 8172 maps of the test set. Therefore, both 2D histograms are build with 8172 values. As in Figure \ref{truevspred}, the red line shows the one-to-one relation while the left histogram shows $\mu$, the measure of the uncertainties on predictions, and the bottom histogram shows $\xi$, a measure of the neural network error.

We first focus on the total mass estimation from the convolutional encoder-decoder. We observe a very good match between the predicted $\mathrm{M_{tot}}$ and the real values for the whole range of total masses ($10^3-10^{4.5}$ M$_\odot$). This tells us that the encoder-decoder is really well suited to predict integrated quantities at all stages of the GC's life, either when the GC is massive and young or when the GC is less massive and older.
As in Fig. \ref{truevspred}, we focus on $\mu$ on the left of each distribution to evaluate the accuracy of our network. In both panels of the figure, we observe a residual curve almost falling on the zero residuals value for the whole range of $\mathrm{M_{tot}}$ and $\mathrm{M_{tot}/L_{tot}}$.
We observe a slight underestimate of the total mass at the highest/lowest mass end, $\mathrm{log_{10}(M_{tot}/M_{\odot})>3.9}$ and $\mathrm{log_{10}(M_{tot}/M_{\odot})<3.4}$. This is because the largest/lowest values of the mass are less frequently encountered while training the network since they constitute the boundary values in the training set.  
The agreement between the predictions and the expectations is also good when looking at the total mass-to-light values (right panel). 
Again, we note some discrepancies at the highest/lowest M/L range.

We report an average percentage error for the total mass estimate of 15.6\% over all the bins, with a minimum value of 6.2\% and a maximum value of 43.5\%. For $\mathrm{M_{tot}/L_{tot}}$, we obtain a mean error of 11.8\%, with a minimum of 5.1\% error and a maximum of 18.1\%.
As for the pixel-by-pixel prediction, we notice again that the largest errors occur for the lowest mass and M/L values. Given the good performance described, we can conclude that the current version of the \jc{encoder-decoder} is already robust in predicting integrated properties with a good accuracy.

\subsubsection{Mass and M/L radial profiles}
\label{massprofandtotmass}

 \begin{figure*}
\begin{center}
    \includegraphics[width=\textwidth]{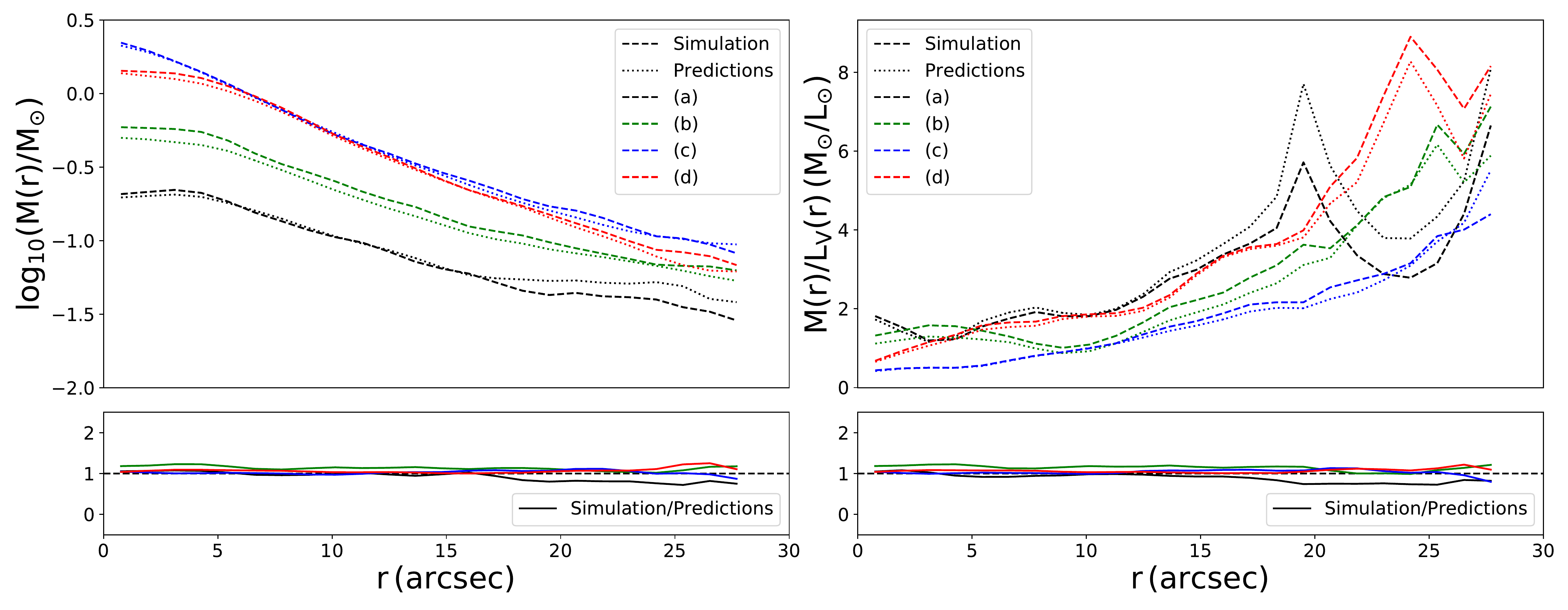}   
\caption{Examples of mass radial profiles (left panel) and M/L profiles (right panel) for four maps from the test set. The different profiles correspond to those of the maps of Fig. \ref{predevery}, labeled as in that figure. The solid lines shows the radial profile in the original simulation while the dashed ones show the profile as predicted by the convolutional encoder-decoder part of \texttt{$\pi$-DOC}. The bottom panel shows the ratio of both curves for the four maps. The predicted radial profiles are in very good agreement with the real ones, and in particular \texttt{$\pi$-DOC} is able to reproduce the shape of the non-constant M/L profiles, indicating mass segregation.}
\label{Rprofile}
\end{center}
\end{figure*}

Finally, we also investigate the mass radial profile of a GC as predicted by the encoder-decoder part of \texttt{$\pi$-DOC}.
In the left panel of Fig. \ref{Rprofile}, we show an example of four mass radial profiles from the test set, corresponding to the four different maps in Fig. \ref{predevery}. The bottom panel of Fig. \ref{Rprofile} shows the ratio between true and predicted curves. The radial mass profile $M(r)$ is computed as the average mass in each radial bin of size $\Delta r$ as follows:

\begin{equation}
 M(r)=\frac{ \sum\limits_{i}^{} {M_i}_{>r-\Delta r/2}^{<r+\Delta r/2} } {\sum\limits_{i}^{} 1}
\end{equation}
where $M_i$ is the mass in each pixel $i$ of a given map.
We choose to use 25 radial bins which translates in $\Delta r \sim 1.2$ arcsecs.

We observe a very good match between the predicted and real mass profile for the four maps of Fig. \ref{predevery}, with an excellent match between the curves in the range 5 to 20 arcsec. This confirms that the network predicts well the shell between the core and the outskirt of the GC at different stage of its life. 
However, depending on the map, we either somewhat under/over predict the mass in the core or in the outskirt of the GC in the range $r<5$ arcsec and $r>25$ arcsec.
This is actually not surprising as we expect the largest/lowest mass values of individual pixels to be found in the inner/outer parts of the maps. As already said in the previous sections, it is an inherent feature of neural network to fail to predict accurately the mass in these ranges compared to the intermediate mass range. We can therefore conclude that the model is already very efficient at predicting the shape of the mass radial profile of a GC never seen during the training phase. 

Moreover, in the right panel of Fig. \ref{Rprofile}, we report the comparison between true and predicted M/L profiles. Notably, the encoder-decoder is able to reproduce the shape of these profiles that are indeed not constant, as contrarily often assumed in standard dynamical models. This is of particular interests because it directly shows that our neural network is able to capture the presence of mass segregation, which causes non-constant M/L profiles as a result of dynamical relaxation processes.

\subsection{Age and distance estimation}
\label{resdistage}

\begin{figure}
\begin{center}
    \includegraphics[width=\columnwidth]{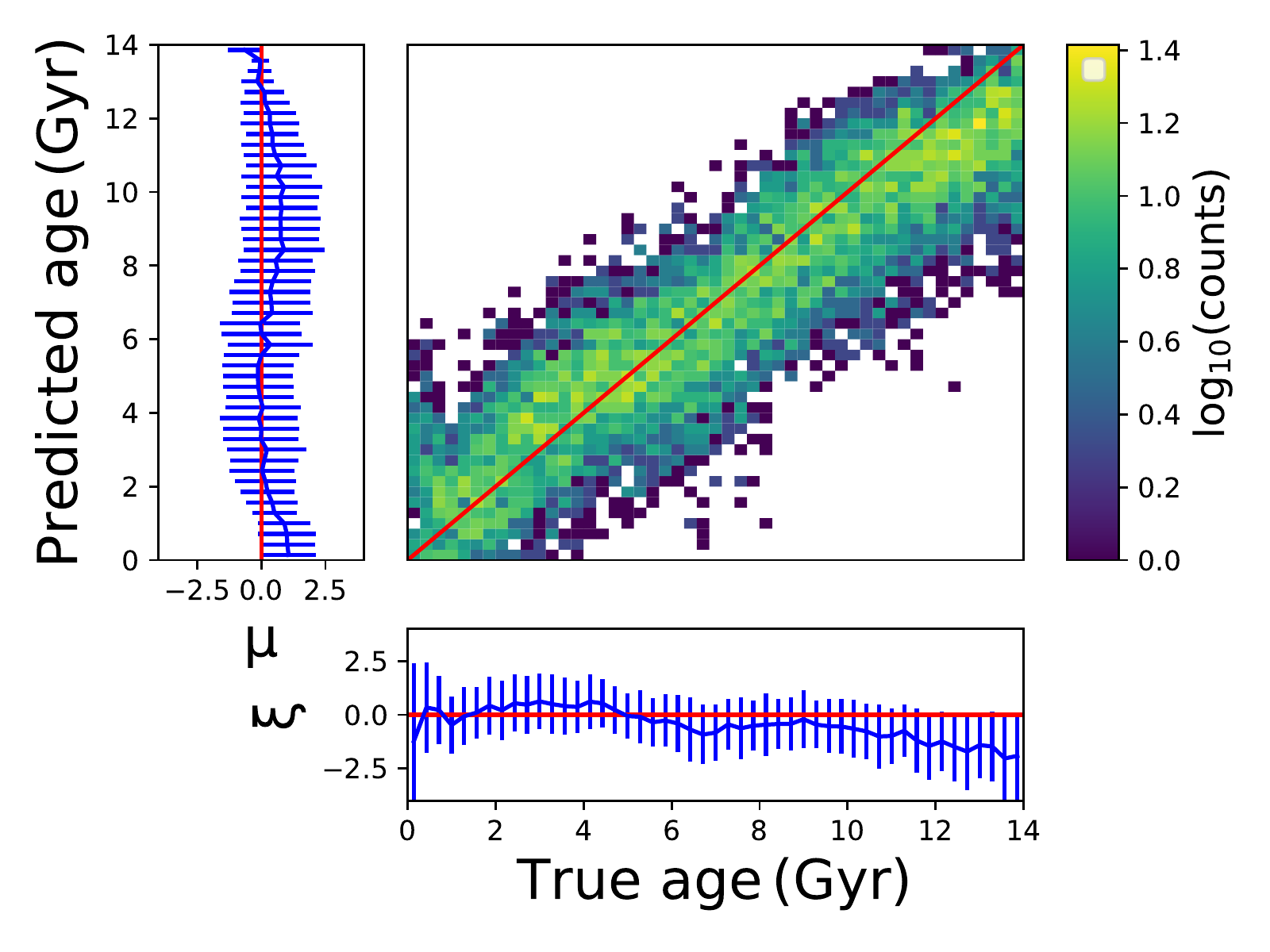}   
\caption{2D distribution of the true versus predicted ages of a GC with the CNN part of \texttt{$\pi$-DOC}. The distribution is calculated over the whole test set with 8172 maps, in which each map corresponds to an age. The red line shows the one-to-one relation and the colour bar indicates the number count in the 2D space of true versus predicted values. The bottom and left histograms show the mean of the residuals $r=\mathrm{age_{predicted}} - \mathrm{age_{true}}$ and associated mean average error. The bottom histogram is the learning error,  p($r$|True), while the side histogram is the recovery uncertainty, p($r$|Predicted).}
\label{true_vs_pred_age}
\end{center}
\end{figure}
 
In this section, we present our results regarding the second part of the \texttt{$\pi$-DOC} architecture: the classic CNN designed to predict the age and distance of a GC from its observed luminosity map.
Such predictions can be seen in Fig. \ref{predevery}, where we indicate the real and predicted ages and distances for the selected four maps. From this figure, we can already see that this neural network predicts values of ages and distances in good agreement with values from the original simulation in a variety of situations (i.e. GCs at different evolutionary stages and at different distances from the observer). We first investigate the performance for the age predictions before looking at the robustness of our distance estimation.  
 
\subsubsection{Age prediction}
\label{ageres}
 
Fig. \ref{true_vs_pred_age} shows the 2D histogram of the true ages versus the predicted values for the 8172 maps of the test set. Each point corresponds to the prediction of the age from an input flux map, resulting in 8172 age predictions in total. At first glance, we report a good agreement between true and predicted values with a distribution falling almost perfectly on top of the one-to-one relation.

We focus on the histogram on the left that shows the distribution of residuals $\mu$ as a function of predictions which encodes the neural network error on the predictions. The average of the residual values are well centred on the zero value. Again, we somewhat overestimate the lowest ages and slightly underestimate the largest ages. This is again due to the inability of neural networks to predict values that are close to the boundary values of the training set.

Conversely to the mass measurements, we estimate the error on our prediction calculating the mean absolute error instead of the mean percentage error in each age:
\begin{equation}
MAE_{\mathrm{bin}} = \frac{1}{\mathrm{N}}\sum{|\mathrm{age}_\mathrm{predicted} - \mathrm{age}_\mathrm{true}|}.
\label{MAE}
\end{equation}
We report an average value of the mean absolute error of 1.26 Gyr over all age bins, with a minimum value of 0.33 Gyr and a maximum value of 1.64 Gyr.
Overall, these values show that our CNN is well suited to predict ages all over the entire life of a GC, starting from a luminosity map only.

 \subsubsection{Distance prediction}
\label{distres}

\begin{figure}
\begin{center}
    \includegraphics[width=\columnwidth]{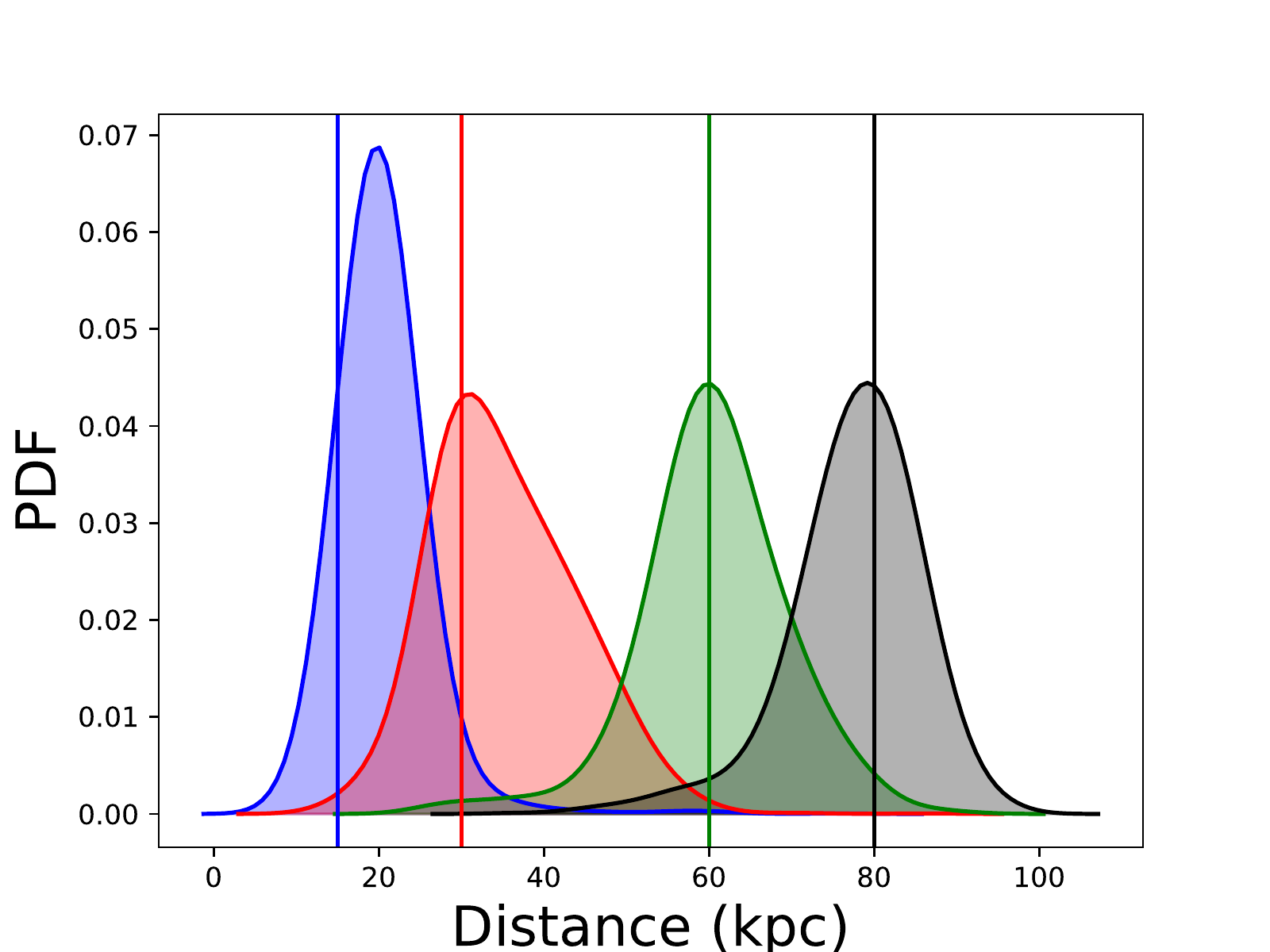}   
\caption{Probability distribution functions (PDF) of the distances predicted with the CNN part of \texttt{$\pi$-DOC}.
Four different PDF are computed for the four real distances constituting the test set (see text for details). The different solid vertical lines show these four distances with colours corresponding to the associated PDF.}
\label{pdf_dist_prediction}
\end{center}
\end{figure}

%--------------------------------------%
\begin{table}

\begin{tabular}{lcccc}    \toprule
Real distance (kpc) &  15 & 30 & 60 & 80 \\\midrule
Mean predicted distance (kpc) &  20.3 & 35.2 & 60.6 & 77.6 \\\midrule
Standard deviation (kpc) &  5.0 & 8.3 & 8.8 & 7.9 \\\midrule
Mean absolute error (kpc) &  5.6 & 7.4 & 6.5 & 6.0 \\
\bottomrule
 \hline
\end{tabular}
 \caption{Summary of the accuracy on CNN predictions for distance estimation.}
 \label{tab2}
\end{table}
%--------------------------------------%

Finally, we investigate how our CNN neural network performs at predicting the distance of a GC from its luminosity map. In Fig. \ref{pdf_dist_prediction} we show the 1D probability function of the predicted distance.
In the test set, we created maps at four different distances (15, 30, 60 ,80 kpc; see Table. \ref{tab1}) which are different from those used in the training set. These four distances are shown on the figure with the different coloured vertical lines. For each map of the test set we know the real distance that should be recovered by the CNN. Therefore, we  build four different PDF for the predicted distances by separating the predicted values according to the real values that should be recovered for each map. 

The PDF of the predicted values  are well centred around the real values for distances of 30, 60 and 80 kpc.
We observe an overestimate of the mean value for the predictions  of GC at a 15 kpc. Moreover, we also report a small underestimate of the mean predicted distance for the 80 kpc case. This is again due to the fact that values close to the boundaries of the training set are more difficult to predict compared to values in the mid-range of the training set. In our case, the training has minimum/maximum values of distances of 10/100 kpc. Therefore the network struggles to predict 80 kpc distances and struggle even more to predict 15 kpc distances since it is closer to the lower boundary.

In Table \ref{tab2}, we summarize the properties of the prediction related to each distribution.
As an example, we observe a very good mean distance estimation for a GC at 60 kpc, with an average value in the distribution of 60.6 kpc and a mean absolute error of 6.5 kpc, corresponding to a $\sim10\%$ relative error. This illustrates that our network is particularly robust for distances in the mid-range of values encountered in the training set. However, we report a mean of the distribution of 20.3 kpc for GCs with real distances of 15 kpc, with a mean absolute error of 5.6 kpc ($\sim$30\% relative error).
Overall, for all of the four distances at test here, we report distance predictions consistent with the true value within one standard deviation. We will discuss in the Section \ref{discusect} the directions we intend to take in order to improve the distance predictions. 

  %%%%%%%%%%%%%%%%%%%%%%%%%%%%%%%%%%%%%%%%%%%%%%%%%%%%%%%%%%%%%%%%%%%%%%%%%%%
 %
 %                      Observations
 %
 %%%%%%%%%%%%%%%%%%%%%%%%%%%%%%%%%%%%%%%%%%%%%%%%%%%%%%%%%%%%%%%%%%%%%%%%%%%%

\section{Testing \texttt{$\pi$-DOC} on observations }
\label{obs}
In this section we test \texttt{$\pi$-DOC}  directly on real observations. This test has to be considered as exploratory, since our algorithm is currently limited by a training set composed of simulations with initial number of particles of 50,000 stars, which is a factor of $\sim10$ smaller than the number of stars in the majority of GCs. This considerably limits the number of GCs for which \texttt{$\pi$-DOC} can be applied on, and, in particular, it excludes the most massive and commonly studied GCs. Based on the \citet{HarrisCat2010} catalog (\citealp{Harris1996},  2010 edition\footnote{\url{http://physwww.mcmaster.ca/~harris/Databases.html}}), we collect a set of GCs with properties in the range of those used in our training: i) total luminosity less than $10^{5}$ L$_\odot$, and  ii) distances between 10 and 100 kpc. We further focus on those GCs for which an homogenous set of observations is available. We use the data from the Pan-STARRS1 Survey (PS1, data release 1; \citealp{Chambers2016}) delivering observations of the sky north of declination $-30^\circ$ in five broadband filters (g, r, i, z, y). We are left with the following 17 GCs, with old ages (>11 Gyr, \citealp{VandenBerg2013}) and distances between $\sim10-40$ kpc, for which we summarize the properties in Table \ref{tab_obs}: NGC~5053, Pal~11, NGC~7492, NGC~5466, NGC~6517, NGC~4590, NGC~6638, NGC~6981, NGC~4147, NGC~7099, NGC~6284, NGC~1904, NGC~5634, NGC~5024, NGC~7006, NGC~5694, NGC~6229.
%--------------------------------------%
\begin{table*}
\begin{tabular}{lccccccccccc}    \toprule
			& L$_\mathrm{FoV}$	& d 			& $R_c$& $R_h$ 		& age 	&[Fe/H] 	& M/L & d & age & M/L & log(M/M$_\odot$)\\
			&[$L_\odot$]&		[kpc]		&  [arcmin]& [arcmin]& [Gyr]&  	&[$M_\odot/L_\odot$]& [kpc]&[Gyr]& [$M_\odot/L_\odot$]&\\	
			& 		& 	(1)	   & 	(1)	& (1)	    &	(2)	& (1)	 & 	(3)				& \texttt{$\pi$-DOC}	&	\texttt{$\pi$-DOC} &\texttt{$\pi$-DOC}	&\texttt{$\pi$-DOC} \\		
\midrule
NGC 5053		&341.23		&17.4 &		2.08  &2.61  	&12.25  	&-2.27	 & $2.60\pm0.59$ &16.8&	13.7&	1.756	&2.78\\ 
Pal 11			&681.17		&13.4		&1.19 & 1.46   	&$-$		&-0.40	&$1.13\pm0.49$ &16.2&	14.9&	0.772	&2.72\\
NGC 7492		&1067.21		&26.3		&0.86 & 1.15   	&$-$		& -1.78	& $1.40\pm0.41$ &21.9&	15.1&	1.022	&3.04\\
NGC 5466		& 1876.25		&16.0		&1.43 & 2.30   	&12.50& 	-1.98	 	& $1.44\pm0.30$ 	&17.3	&8.5	&1.217	&2.87\\	
NGC 6517		&2026.29		&10.6&		0.06  &0.50   	&$-$&	-1.23 	& $2.27\pm0.67$ &30.1&	4.5&		6.995	&4.15\\	
NGC 4590		&5224.34		&10.3&	 	0.58  &1.51  	&12.00 & 	-2.23		& $2.01\pm0.22$ &21.4&	2.1&		1.365	&3.85 \\ 	
NGC 6638		&7366.47		&9.4	&		0.22  &0.51   	&$-$& 	-0.95 	& $1.34\pm0.35$ &12.4& 0.4&		1.755	&4.11\\
NGC 6981		&9888.63		&17.0&		0.46  &0.93   	&11.50&	-1.42 	& $1.60\pm0.31$ &25.6&	1.2&		0.779	&3.89\\
NGC 4147		&10269.37	&19.3&		0.09  &0.48   	&12.25 &	-1.80	 	& $1.66\pm0.41$ &27.6&	0.4&		1.190	&4.09 \\	
NGC 7099		&11351.31	&8.1	 &		0.06  &1.03   	&13.00&	-2.27 	& $2.04\pm0.17$ &17.4&	-1.9&	1.094	&4.09\\	
NGC 6284		&16612.68	&15.3&		0.07 & 0.66   	&$-$&	-1.26 	& $1.53\pm0.39$ &16.9&	0.2&		0.884	&4.17\\
NGC 1904		&25315.31	&12.9&		0.16  &0.65   	&$-$&	-1.60	 	& $1.47\pm0.15$ &17.0&	-1.0&	0.513	&4.11\\	
NGC 5634		&32111.04	 &25.2&		0.09  &0.86   	&$-$&	-1.88	 	& $1.91\pm0.46$ &26.7&	1.8&		0.412	&4.12\\
NGC 5024		&37269.46	&17.9&		0.35  &1.31  	&12.25 &	-2.10		 & $1.74\pm0.17$ &16.8&	-0.5&	0.241	&3.95 \\	
NGC 7006		&42293.77	&41.2&		0.17  &0.44   	&11.25	& -1.52	& $1.52\pm0.40$ &37.9&	3.5&		0.320	&4.13\\
NGC 5694		&50432.94	&35.0 &		0.06 & 0.40   	&$-$		&-1.98	 & $2.08\pm0.42$ &27.3&	1.9&		0.272	&4.1 \\	
NGC 6229		&68421.00	&30.5&		0.12 & 0.36  	&$-$		&-1.47	 & $1.89\pm0.63$  &20.3&0.4&		0.220	&4.18\\

\bottomrule
 \hline
\end{tabular}
 \caption{Properties of the GCs on which we apply \texttt{$\pi$-DOC} using the Pan-STARRS1 Survey data. We report the V-band luminosity in the observed FoV, the distance to the cluster, its core and half-light radius, the age, the metallicity, and measured M/L ratio. Distance, age, M/L and mass in the FoV predicted by \texttt{$\pi$-DOC} are also reported. The literature values of the clusters are from: (1) \citet{HarrisCat2010} catalog; (2) \citet{VandenBerg2013}; (3) \citet{Baumgardt2020}}
 \label{tab_obs}
\end{table*}
%--------------------------------------%

\begin{figure}
\begin{center}
    \includegraphics[width=0.45\textwidth]{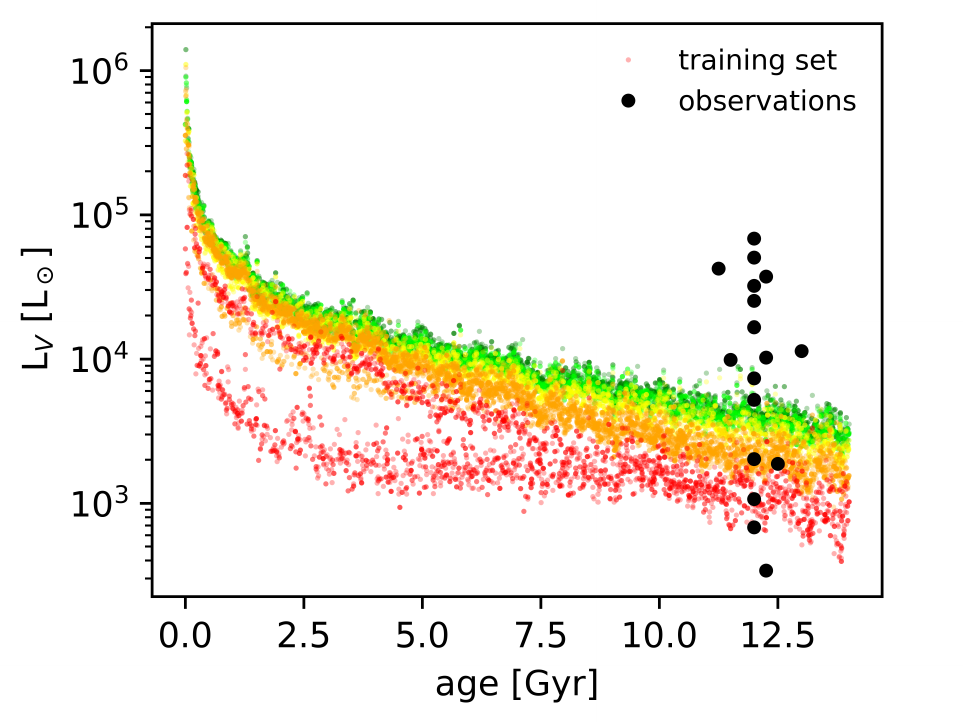}   
\caption{V-band luminosity enclosed in the FoV vs. age of the snapshots employed in the training set (coloured points) and of the observed GCs (black points). Different colours for the training set refer to the different distances used (10 kpc to 100 kpc, in red, orange, yellow and green). Out of the 17 GCs selected for observations, 5 low-luminosity GCs fall within the luminosity limit of the training set, given their ages and distances (>11 Gyr and 10-40 kpc). We expect \texttt{$\pi$-DOC} to perform satisfactorily for these 5 GCs (NGC~5053, Pal~11, NGC~7492,  NGC~5466 and  NGC~6517).}
\label{comparison_train}
\end{center}
\end{figure}

PS1 has a pixels scale of 0.25 arcsec and a median seeing between 1 and 1.3 arcsec, consistent with the pixel scale and seeing conditions used in our algorithm. To obtain the observed maps, we select a FoV of 40x40 arcsec$^2$ around the GCs centres from PS1 stacked images (\citealp{Waters2020}), and for each cluster, the photometry in g- and r-bands is transformed into V-band photometry using the empirical relation described in \citet{Kostov2018}.\footnote{ This transformation is based on the calibration of the PS1 photometric system and Stetson's BVRI standard star catalogue (\citealp{Stetson2000}).} Finally, following the procedure used for the mock observations, we smooth the GCs flux maps.
Before applying \texttt{$\pi$-DOC} on these maps, we estimate the total luminosity enclosed in the FoVs, using the distance from the \citet{HarrisCat2010} catalog (see Table \ref{tab_obs}).
This allows us to directly compare the FoV luminosities employed in the training set and those observed, as a way to preventively assess for which GCs we could expect to obtain reliable predictions. In Figure \ref{comparison_train} we show this comparison highlighting the dependencies of our training set on age and distance. In the age and distance range of our observations (>11 Gyr and 10-40 kpc), the luminosity of the training set is in the range of $\approx300-5000$ L$_\odot$, with an average of 1754 L$_\odot$. Only 5 clusters in our sample match this luminosity regime, namely NGC~5053, Pal~11, NGC~7492, NGC~5466, and NGC~6517. For these clusters we expect \texttt{$\pi$-DOC} to give satisfactory predictions. The other GCs can instead be used to test the ability of \texttt{$\pi$-DOC} to generalize on data outside of the training set limits and to point out the limitation of the current training set.

\begin{figure*}
\begin{center}
    \includegraphics[width=\textwidth]{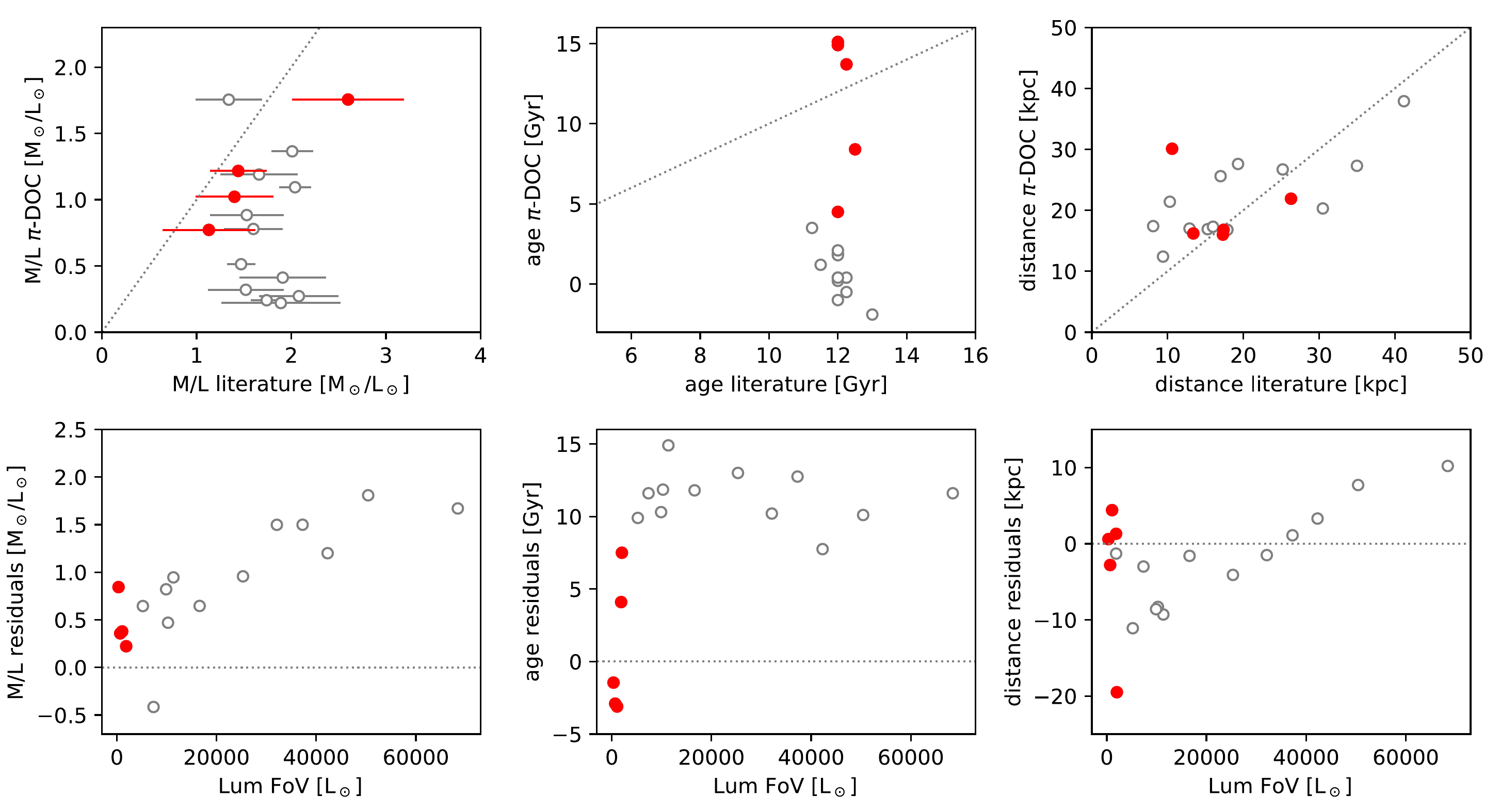}   
\caption{\textit{Top panels:} Predicted M/L, age and distance vs. literature values for the 17 GCs analyzed; the dashed lines represent the one-to-one relations. \textit{Bottom panels:} Residuals between literature and predicted M/L, age and distance as a function of the luminosity enclosed in the FoV. The 5 GCs with the least luminous FoVs are reported as red points and represent those GCs for which the algorithm \texttt{$\pi$-DOC} is expected to better perform.}
\label{fig_obs}
\end{center}
\end{figure*}

\begin{figure*}
\begin{center}
    \includegraphics[width=\textwidth]{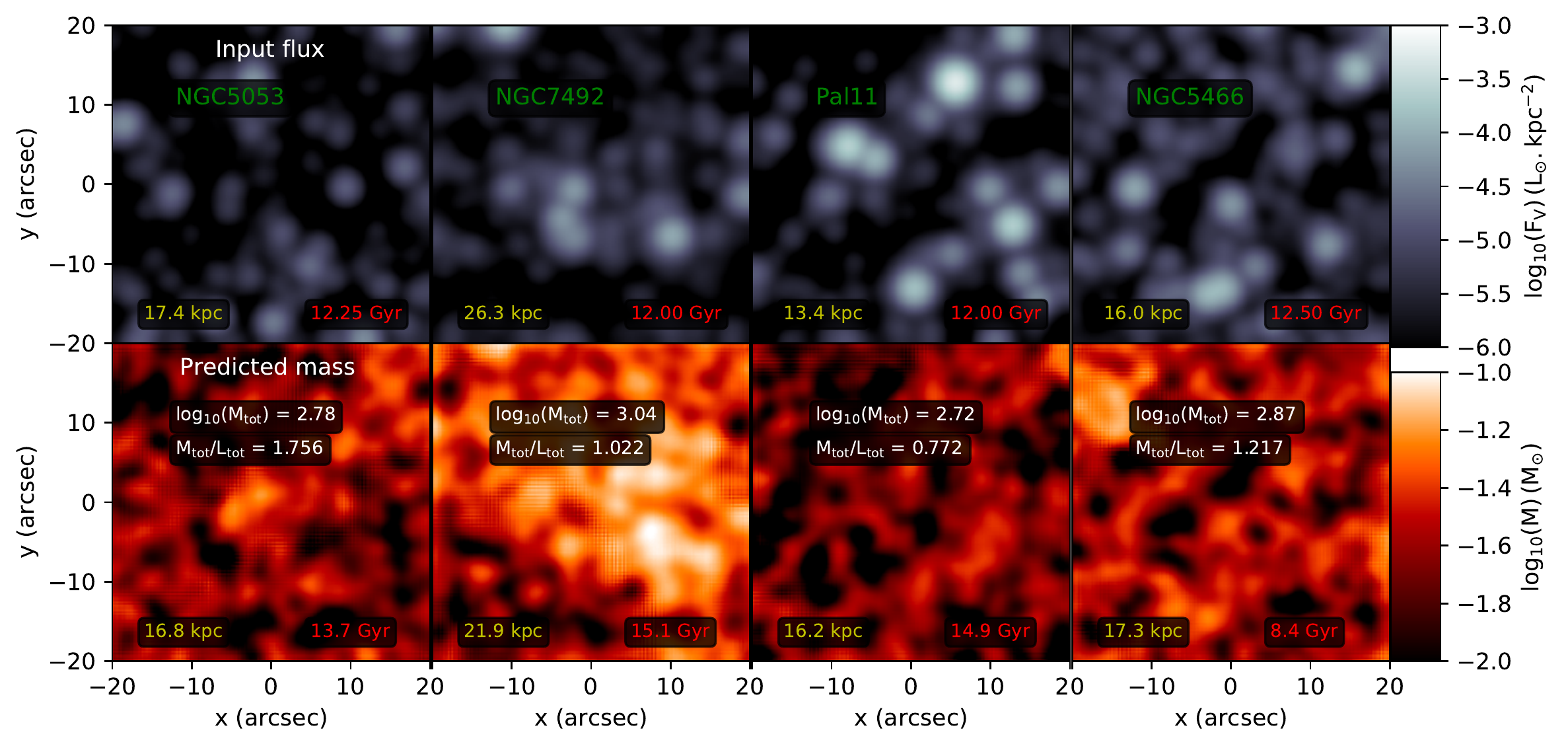}   
\caption{Predictions of \texttt{$\pi$-DOC} for the 4 least luminous GCs (NGC~5053, Pal~11, NGC~7492, NGC~5466) for which M/L, age and distance values consistent with the literature are retrieved. The first row reports the input V-band flux from PS1 observations, with associated age and distances (from Table \ref{tab_obs}). The second row reports the mass maps obtained by \texttt{$\pi$-DOC}, the associated mass and M/L in the FoV and the age and distance predictions.}
\label{maps_obs}
\end{center}
\end{figure*}

In order to compare the results of \texttt{$\pi$-DOC} with the literature, we collect the distance from the \citet{HarrisCat2010} catalog, the age from \citealp{VandenBerg2013}\footnote{If an age estimate is not available we assume an age of 12 Gyr.} and the global M/L in the V-band from \citet{Baumgardt2020}. Notice that mass profiles are not available for our selected sample of GCs, but there are only global estimates of the total mass.\footnote{Mass profiles only exists for a handful of clusters, namely the most massive and extensively studied GCs.}  These total masses are not suitable for a direct comparison with our results, since the current version of \texttt{$\pi$-DOC} gives the mass enclosed in a relatively small FoV (with a size smaller than the typical half-light radius, see Table \ref{tab_obs}). We therefore concentrate on a comparison with the M/L literature values which only weakly depend on the size of the FoV. However, as seen in Section \ref{massprofandtotmass}, we stress that the M/L profile is expected to be non-constant due to mass segregation, which induces lower values of M/L in the centre, due to the preferential presence of massive, low-M/L, stars. Therefore, at this stage, the M/L values obtained by \texttt{$\pi$-DOC} could be considered as central values, possibly underestimating the global M/L from the literature.

We report the results from \texttt{$\pi$-DOC} in Table \ref{tab_obs} and, in Figure \ref{fig_obs}, we show the comparison between the predictions and literature values, for the M/L, age and distance of the 17 GCs. In the same figure, we show the residuals between literature and predicted values as a function of the GCs luminosity enclosed in the FoV. Moreover, in these plots, we indicate with red symbols those five low-luminosity GCs for which our algorithm is expected to work. For these five GCs, \texttt{$\pi$-DOC} is able to obtain M/L values fully consistent with the literature for 4 out of 5 GCs: NGC~5053, Pal~11, NGC~7492 and NGC~5466. The corresponding maps are presented in Figure \ref{maps_obs}. For these 4 clusters the distance estimate is also fully consistent with the literature, and for the three least massive GCs, we are also able to obtain consistent age estimates. We notice that for NGC~6517, we obtain a M/L value of M/L=6.995 $M_\odot/L_\odot$, a factor of $\approx$3 times larger than the expected values. Interestingly, also the distance estimate is a factor of 3 higher, therefore if we rescaled the luminosity accordingly, we would obtain a value of M/L consistent with the literature.

From Figure \ref{fig_obs}, it is evident that \texttt{$\pi$-DOC} performs well within the limits of the training set and, as a consequence, for the GCs with high FoV luminosities we are not able to obtain values consistent with the literature. For these clusters, the algorithm interprets their high-luminosity as a sign of young age (at a fixed number of stars, young clusters are more luminous because of stellar evolution) and, as a result, the M/L values are severely underestimated. The trend of worse performances with higher luminosity is confirmed by the M/L residuals plot in Figure \ref{fig_obs}, in which the residuals correlate with the FoV luminosity. This clearly points out the limit of the current training set of \texttt{$\pi$-DOC} and highlights the need for larger N-body simulations. Finally, we notice that the distance estimates, with the exception of NGC~6517, are consistent even for those clusters with luminosity outside the range of the training set. However, a large scatter remains, as already discussed in Section \ref{distres}.

 %%%%%%%%%%%%%%%%%%%%%%%%%%%%%%%%%%%%%%%%%%%%%%%%%%%%%%%%%%%%%%%%%%%%%%%%%%%
 %
 %                      Discussion/conclusion section 5
 %
 %%%%%%%%%%%%%%%%%%%%%%%%%%%%%%%%%%%%%%%%%%%%%%%%%%%%%%%%%%%%%%%%%%%%%%%%%%%%

% \section{Discussion and Conclusion}
 %\label{discussion}
  
%In this section, we discuss the global results of our deep learning framework \texttt{$\pi$-DOC} with its current success and current limitations.
%We finally give our conclusion and what we expect for the near future to improve our methodology.

\section{Discussion and conclusions}
\label{discusect}

In this work we have developed a new method to predict the dynamical mass distribution, the age, and the distance of a GC, starting from its luminosity map only, and leveraging on the application of machine learning techniques in combination with dynamical $N$-body simulations of GCs. The algorithm, \textit{Predicting Images for the Dynamics Of stellar Clusters}, \texttt{$\pi$-DOC}, consists of a combination of two neural networks (a convolutional encoder-decoder and a convolutional neural network) trained on $N$-body simulations of GCs. The algorithm is capable of learning the underlying physics involved in the transformation between flux maps and the intrinsic mass distribution in the given FoV and, simultaneously, it predicts for each map the distance and the age of a GC.

Our technique introduces several advantages:
\begin{itemize}
\item The dynamical effects that play a fundamental role in shaping the mass distribution of a GC (e.g. mass segregation, anisotropy, mass loss) are directly taken into consideration without the need of developing complex multi-component dynamical models, since \texttt{$\pi$-DOC} is trained on direct $N$-body simulations that \textit{naturally include all dynamical and evolutionary effects}.
\item \textit{Observational biases are directly taken into consideration} in our forward-modelling approach, since the algorithm is trained on mock observations derived from simulations, which include, for example, seeing effects and a large range of distances. This makes \texttt{$\pi$-DOC} a very flexible tool potentially applicable to a variety of GCs observations (both nearby and far).
\item The prediction of a dynamical mass map is \textit{extremely fast} and only takes about 70 milliseconds.
\item The growing number of direct $N$-body simulations being developed by the community makes it \textit{feasible to extend such an algorithm} in the future, in order to sample more realistic and comprehensive ranges in the parameter space.
\end{itemize}

The current version of the algorithm has been trained on a limited set of simulations with 50,000 stars, which translates in FoVs with dynamical masses of $10^3-10^{4.5}$ M$_\odot$. Despite the low-number of particles, these simulations provide us with a standard description of the long-term evolution of a GC and its dependency on initial conditions (e.g. different initial half-mass radii and densities). We specifically tested our algorithm on a test set taken from a simulation never seen during the training and treated as a mock observation. The test set includes intermediate initial density, distances and seeing conditions not employed for the training. As a result, we demonstrated that \texttt{$\pi$-DOC} is able to generalize over a large range of parameters never seen during the training. This is a very important feature for planning a future extension of the training set and for future applications to observations.

The performance of the mass-prediction part of the algorithm, the convolutional encoder-decoder, is very satisfactory: \texttt{$\pi$-DOC} is able to predict the mass pixel-by-pixel with a mean error of $\sim$27.4\% within the entire range sampled, and the total mass in a FoV with an error of $\sim$15.6\%. In both cases the maximum percentage errors occur in the low-value regimes of the sampled ranges, at $\mathrm{log_{10}(M/M_{\odot})<-1.5}$ and $\mathrm{log_{10}(M_{tot}/M_{\odot})<3.2}$, respectively. The mean performance excluding these low values is significantly better (e.g. mean error of 17.6\% for the pixel-by-pixel mass prediction).

As a direct consequence, we are also able to recover the total M/L for each map with an average error of $\sim$11.8\%, without any significant systematics. This makes \texttt{$\pi$-DOC} a competitive tool that can offer the prospect of performing accurate measurements of dynamical M/L, which are today still missing (\citealp{Sollima2015,Zocchi2017,Bellini2017}). Moreover, we demonstrated that we are able to obtain accurate predictions for the mass profiles (and associated M/L profiles), which allow one to study the spatial distribution of stars (and possibly stellar remnants) due to mass segregation, the direct consequence of dynamical relaxation. Importantly, we stress that the accuracy of all our mass predictions does not depend on neither the age, distance nor evolutionary stage of the cluster.

Furthermore, the CNN part of \texttt{$\pi$-DOC} is able to correctly predict age and distance with a mean error of $\sim1.5$ Gyr and $\sim6$ kpc, respectively. The performance in this case is not yet competitive with other standard techniques and it is likely due to the fact that we only use photometric information in a single band (the V-band in our case). Moreover, concerning the distance, we notice that the algorithm is trained on a set of only 5 distance values, which is likely the main limiting factor for accurate predictions. Despite this limitation, \texttt{$\pi$-DOC} is able to generalize for the distances never seen in the training, as reported in Figure \ref{pdf_dist_prediction}. A simple increase of the number of distances employed in the training would improve the performances.

A general drawback that we highlighted for both the encoder-decoder and the CNN part of the algorithm is the inability of the neural network to predict values too close of the boundary values employed in the training set. This is an inherent problem of neural network training experiments extensively discussed in the literature (see \citealt{2019MNRAS.484..282G} and \citealt{2019MNRAS.490.1055C} for example), and it is due to the fact that a trained neural network cannot predict values that overshoot or undershoot values from the training set. Therefore, this generates a sort of pile-up of the predictions just below the maximum value and just above the minimum value from the training set. One way to palliate this issue is to include a larger range of values in the training set and use the final algorithm only within a restricted parameter range that excludes the boundaries.

As discussed above, the performances of \texttt{$\pi$-DOC} tested on mock observations are successful and demonstrate that the neural network architecture used here is well suited for the task of measuring dynamical mass, age and distance of a cluster. 
As a further test, we applied \texttt{$\pi$-DOC} to a set of observations with the goal of both analyzing the actual performance on real data of the current version of the algorithm and highlight the points that still need improvement. We selected 17 low-luminosity GCs in order to match as much as possible the range of the physical properties included in the training set. Using the Pan-STARRS1 Survey, we extracted for each of the clusters the associated flux maps and applied our algorithm on them. \texttt{$\pi$-DOC} is able to simultaneously recover M/L, age and distance consistent with the ones reported in the literature only for the GCs with the lowest luminosities enclosed in the FoV. This is explained by the fact that higher FoV luminosities are not included in the training and therefore the algorithm fails at predicting correctly the mass and the age. However, we notice that \texttt{$\pi$-DOC} is able to recover satisfactory distance values also for those GCs with high FoV luminosities.

Our analysis demonstrates that \texttt{$\pi$-DOC} performs well within the limits of the training set, that for the moment only includes small GCs simulations. Before a systematic application to a large set of observations, a series of improvements to both the training set and the code structure are needed, as highlighted by the limitations that emerged in this work. For this reason we summarize here the main future developments that we anticipate to address in a followup work:
\begin{itemize}
\item Extension of the training set, including more comprehensive simulations with high number of stars (up to $10^6$), a variety of strengths of the tidal field, and different metallicities, as a way to reproduce densities and mass scales typical of observed GCs.
\item Select a setup of the FoV (e.g. FoV size, pixel scale) that is adaptable to a wide set of observational conditions, possibly including the features of both galactic and extragalactic GC observations.
\item Include a photometric colour information as a way to improve age and distance predictions. This could be done, for example, by using as an input a luminosity map in the B-band simultaneously to the the V-band map.
\item Include kinematic information, for example the velocity dispersion in a given FoV. Since the mass is directly proportional to the square of the velocity dispersion (virial theorem), accounting for kinematics in the input will improve the performance of the dynamical mass prediction and even open up the possibility of predicting the presence of non-visible mass, with possible application to dark-matter dominated systems and systems containing dark remnants.
\item Consider the possible presence of background contamination, which could have a non-negligible effect especially for low-density GCs situated in high density regions.
\end{itemize}

In this work we demonstrated that the current neural network architecture of \texttt{$\pi$-DOC} is suitable for learning the physics underlying $N$-body dynamics occurring in the evolution of GCs, in particular for learning the transformation between dynamical mass and luminosity. Our encouraging results, tested both on mock observations and real data, indicate that \texttt{$\pi$-DOC} can represent a viable new and fast approach for the determination of GCs properties, extremely competitive and independent of the classical dynamical modelling techniques.

%\jc{However, the current version of our model has already shown a large variety of success. We have actually shown the ability of our model to generalize on a large range of parameter in the parameter space investigated here. First we used two values of the seeing conditions (PSF) to mimic real observations with actual telescope inside the training set while we tested the prediction with a PSF in between. It appears that the model learnt well the relation between mass, age and distance estimation with PSF as the model extrapolates well when seeing luminosity maps with a different PSF.}  

\section*{Acknowledgments}
We thank the referee for their comments and suggestions.
The authors were granted access to the HPC resources of CINES and IDRIS under the allocation A0070411049 attributed by GENCI (Grand Equipement National de Calcul Intensif) and the Jean-Zay Grand Challenge (CT4) ``Emulation de simulations de R\'eionisation par apprentissage profond''. 
We thank Alison Sills and Meghan Miholics for providing access to the $N$-body simulations and Oliver M\"uller for useful discussions.
We thank contributors to SciPy, Matplotlib, and the PYTHON programming language.
We thank the Keras and Talos API for deep learning machinery and optimization in PYTHON. 

The Pan-STARRS1 Surveys (PS1) and the PS1 public science archive have been made possible through contributions by the Institute for Astronomy, the University of Hawaii, the Pan-STARRS Project Office, the Max-Planck Society and its participating institutes, the Max Planck Institute for Astronomy, Heidelberg and the Max Planck Institute for Extraterrestrial Physics, Garching, The Johns Hopkins University, Durham University, the University of Edinburgh, the Queen's University Belfast, the Harvard-Smithsonian Center for Astrophysics, the Las Cumbres Observatory Global Telescope Network Incorporated, the National Central University of Taiwan, the Space Telescope Science Institute, the National Aeronautics and Space Administration under Grant No. NNX08AR22G issued through the Planetary Science Division of the NASA Science Mission Directorate, the National Science Foundation Grant No. AST-1238877, the University of Maryland, Eotvos Lorand University (ELTE), the Los Alamos National Laboratory, and the Gordon and Betty Moore Foundation.

We thank Aur\'elian Chardin for giving us the idea for the name of the algorithm.

\section*{DATA AVAILABILITY}
The data underlying this article will be shared on reasonable request to the corresponding author.

% BIB =========================================================================================
\bibliographystyle{mnras}
\bibliography{biblio}

\appendix

\section{Perceptual loss function for the convolutional encoder-decoder}
\label{perceploss}

Here we detail the loss function used in the convolutional encoder-decoder neural network part of \texttt{$\pi$-DOC} for the spatial mass distribution predictions.
When training neural networks it is a common feature to use the mean square error loss function between ground-truth and predictions.
Some more exotic loss functions, such as the mean absolute error, the binary crossentropy (for binary classification tasks), categorical crossentropy (for multiple categories classification tasks) or even more complicated functions, are also used depending on the deep learning problem we are facing. Overall, such loss functions aim at calculating a single number evaluated from the current prediction of the model and the actual ground truth we want to recover. Then the deep learning machinery aims at minimizing this number for all examples in the training set thanks to the optimizer. 

Such loss functions are known to work well for scalar predictions such as parameter estimations.
This is why we are using the mean square error for the convolutional neural network part of \texttt{$\pi$-DOC} when we aim to predict age and distances of GCs.
However, when we try to train an encoder-decoder, such a number will be calculated over the whole maps
and we are left with a per-pixel loss function that does not guarantee to capture the perceptual differences between predictions and ground-truth images. For example, two identical images offset from each other by only one pixel have a high degree of perceptual similarity but could have a very different per-pixel loss.  
In order to focus more on spatial pattern recovery, perceptual loss functions have been introduced to compare two different images that look similar.

\citet{Johnson2016Perceptual} among others have shown that perceptual loss functions are more accurate in generating high quality images.
In our case we build a perceptual loss function in the exact same spirit of \citet{Johnson2016Perceptual} who were inspired by \citet{Simonyan2013}, \citet{Szegedy2014}, \citet{Yosinski2015} and \citet{Nguyen2016}.
The main idea behind such methodology is to use a convolutional neural network pre-trained for image classification that has already learned to encode perceptual informations. In our case we use the well known VGG16 neural network developed by \citet{Simonyanvgg16}.
VGG16 is an image classification convolutional neural network that achieves 92.7\% accuracy in ImageNet, which is a dataset of over 14 million images belonging to 1000 classes.

The methodology is then to feed this network one time with the current output image of our encoder-decoder and one time with the ground truth image we aim to predict. Then, we take the output image of VGG16 in a convolutional layer of the network that we chose for both images.
We are then left with two images and we compute the mean square error between these two images.
In practice we can compute a mean square error value for each of the outputed images in the different convolution layers in VGG16.
We can then build a perceptual loss function which is the sum of all these mean square values. 
Because the output images at the end of each convolutional layer enlight different spatial informations, we then add informations in the final loss function compared to a single per-pixel loss functions. In our case, we add the mean square error between the true output and predicted output of three convolutional layers in VGG16 which have the following names : block1\_conv1, block1\_conv2 and block2\_conv2.

\end{document}